\documentclass[journal]{IEEEtran}
\usepackage[left=0.75in, right=0.75in, top=1in, bottom=0.9in]{geometry}
\makeatletter
\def\ps@headings{%
\def\@oddhead{\mbox{}\scriptsize\rightmark \hfil \thepage}%
\def\@evenhead{\scriptsize\thepage \hfil \leftmark\mbox{}}%
\def\@oddfoot{}%
\def\@evenfoot{}}
\makeatother 

\usepackage{amsmath,amssymb,mathrsfs,amsthm,,bm,bbm}
\usepackage{mathtools}
\usepackage{cite}

\usepackage{graphicx,stfloats,subfigure,multicol}
\usepackage{epsfig,epsf,psfrag,amssymb,amsfonts,latexsym,graphicx,mathrsfs,subfigure}

\usepackage{comment}

\usepackage{booktabs}
\usepackage[table]{xcolor}
\usepackage{multirow}
\usepackage{tablefootnote}
\usepackage{bbm}
\usepackage{chngpage}
\usepackage{textcomp}
\usepackage{hyperref}
\usepackage{url}
\usepackage[hyphenbreaks]{breakurl}
\usepackage[normalem]{ulem}
\usepackage{algpseudocode}
\newsavebox{\ieeealgbox}

\usepackage{array}

\newtheorem{theorem}{Theorem}

\newtheorem{lemma}{Lemma}

\usepackage{comment}
\usepackage{algorithm,algpseudocode}
\usepackage{url}

 \def\old#1{}

\def\nn{\nonumber}
\def\beq{\begin{equation}}
\def\eeq{\end{equation}}
\def\bea{\begin{eqnarray}}
\def\eea{\end{eqnarray}}
\def\ba{\begin{array}}
\def\ea{\end{array}}

\def\bitem{\begin{itemize}}
\def\eitem{\end{itemize}}
\def\ben{\begin{enumerate}}
\def\een{\end{enumerate}}

\definecolor{bgrd}{rgb}{1,1,1}
\definecolor{gray}{rgb}{0.5,0.5,0.5}
\definecolor{dkr}{rgb}{0.7,0.1,0.2}
\definecolor{dkb}{rgb}{0.1,0.1,0.8}

\def\tcb{\textcolor{blue}}

\begin{document}

\title{Integrating Distributed Energy Resources: Optimal Prosumer Decisions and Impacts of Net Metering Tariffs}

\author{Ahmed S. Alahmed and
Lang~Tong
\thanks{\scriptsize  Ahmed S. Alahmed and
Lang Tong  ({\tt \{\tcb{ASA278,~LT35}\}\tcb{@cornell.edu}}) are  with the School of Electrical and Computer Engineering, Cornell University, Ithaca, USA.}}
\maketitle

{
\begin{abstract}
The rapid growth of the behind-the-meter (BTM) distributed generation has led to initiatives to reform the net energy metering (NEM) policies to address pressing concerns of rising electricity bills, fairness of cost allocation, and the long-term growth of distributed energy resources.  This article presents an analytical framework for the optimal prosumer consumption decision using an inclusive NEM X tariff model that covers existing and proposed NEM tariff designs.  The structure of the optimal consumption policy lends itself to near closed-form optimal solutions suitable for practical energy management systems that are responsive to stochastic BTM generation and dynamic pricing.  The short and long-run performance of NEM and feed-in tariffs (FiT) are considered under a sequential rate-setting decision process. Also presented are numerical results that characterize social welfare distributions, cross-subsidies, and long-run solar adoption performance for selected NEM and FiT policy designs.

\end{abstract}
}

\section{Introduction}
\label{sec:intro}
Much of the recent debate on retail electricity tariff centers around the net energy metering (NEM)  policies that played a critical role in the phenomenal growth of the behind-the-meter (BTM) distributed generation, mostly from rooftop photovoltaics (PV).  First implemented serendipitously in the late 1970s without the awareness of the utility company\footnote{\url{https://en.wikipedia.org/wiki/Net_metering}}, the classical NEM tariff, commonly referred to as NEM 1.0, offers compelling economic incentives for PV adoption by compensating a {\em prosumer\footnote{We call a customer capable of BTM generation a {\em prosumer} and one without such capability a {\em consumer.}}} for its export of net BTM generation at the same retail price as for consumption.

It turns out that a kilowatt-hour (kWh) production of BTM PV is worth differently to different participants in the retail electricity market.  To prosumers, every kWh of energy from their BTM PV  is worth what the utility charges them for consumption.  To a regulated utility company, a kWh exported by prosumers costs far more than what the utility can buy from the wholesale electricity market, primarily due to cost bundling in retail tariff design. The loss of revenue due to BTM generation and the non-economic payment to prosumers' export cause the increase in the retail price of electricity. To consumers without BTM PV, that kWh production from BTM PV for prosumers means the shifting of the grid operation cost from prosumers to consumers, resulting in a cross-subsidy of prosumers by consumers.  With many consumers lacking the means of BTM PV investments, such cost-shifts raise the normative question of fairness, which the state regulator must address in its rate-setting process.

By 2021, almost all of the 50 states in the US have begun considering reforms to their existing NEM tariff models, with early adoptions of NEM 2.0 policies in multiple states.    Discussions on the implementation of NEM 2.0 and its successor NEM 3.0 have generated sometimes contentious debates about the costs and economic opportunities of solar.  A major change of NEM 1.0 in the NEM 2.0 and NEM 3.0 policies is the lowered compensation rate for net production.  Other changes under discussion include discriminative pricing that separates consumers and prosumers.

The proposed policy changes have significant economic and engineering implications.  For instance, the differentiated pricing of net consumption and production implies that the BTM production is valued differently, depending on whether it is consumed locally or exported to the grid.   When the time-of-use (TOU) features are incorporated, the time of PV production becomes a factor. In response to these changes, the engineering design of energy management systems must optimize the allocation of the BTM generation to the set of demands and the possibility of exporting the generation to the grid.

We present an analytical framework centered around prosumers' optimal consumption decisions under an inclusive tariff model covering all existing and most proposed NEM policies.  With ongoing policy debates on the evolving NEM proposals, this work aims to bridge the gap between the engineering energy management design and the economic implications of NEM tariff choices.

The rest of the paper is organized as follows: Sec.~\ref{sec:tariffmodels} presents NEM-X---a general NEM tariff model covering most of the implemented and proposed NEM tariffs.  The prosumer decision problem is analyzed in Sec.~\ref{sec:prosumer} where we consider the problem of optimal consumption of {\em active prosumers} who set consumption levels based on available BTM DER generations.   The two-threshold structure of the optimal consumption policy is characterized, which gives a near closed-form solution for the optimal consumption bundle.  In dealing with stochasticities of the BTM DER generation, a model-predictive scheduling strategy is proposed, leveraging the structure of the optimal consumption policy.  Prosumer surplus expressions are obtained and compared under NEM and feed-in tariff (FiT) policies for active and {\em passive prosumers} whose consumption is independent of BTM production and who use all DER generations for bill saving.  Sec.~ \ref{sec:RegulatorProblem} presents a system theoretical model for the regulator's rate-setting problem where NEM tariffs are endogenously determined based on the DER adoption level and utility's anticipated break-even costs.  Measures of welfare distributions, cross-subsidies, and market potential are presented in Sec.~\ref{sec:adoptionModelDynamics}, and numerical results are presented in Sec.~\ref{sec:numerical}.

\par The literature on NEM and FiT policies is quite extensive.  A contextual survey is provided in Sec.~\ref{sec:appendixLR}, focusing on social welfare distribution, cross-subsidies, and long-run performance when the NEM and FiT rates are set sequentially.
The NEM X tariff model was first proposed in \cite{Alahmed_Tong:22IEEETSG}, which generalizes earlier models of NEM, FiT, and net purchase and sale tariffs by the work in \cite{YamamotoPricingEF:12SE}.   There is little published work on the optimal prosumer decision problems under NEM X.  The results presented here are built on \cite{Alahmed_Tong:22IEEETSG} with several extensions addressing the stochasticity of BTM renewables and long-run performance of NEM X.  Most theoretical and algorithmic details are relegated to Appendix \ref{sec:appendixProofs}. Appendix \ref{sec:appendixNotations} provides a list of major notations and symbols. Extra numerical results are provided in Appendix \ref{sec:appendixNotations}, and the data used for implementing the empirical study are further detailed in Appendix \ref{sec:appendixNumData}.

\section{Net metering and feed-in Tariff Models}
\label{sec:tariffmodels}
 We consider a retail electricity market consisting of a regulated utility company, consumers that do not have BTM DER,  and prosumers with BTM DER. The regulator\footnote{Typically the public utilities commission.} sets the rates of consumption and production that define the retail tariff governing the customer payments. Most, if not all, retail tariffs belong to either the NEM or the FiT tariff families; the U.S. markets have mostly adopted the variants of the NEM tariff, whereas FiT is more prevalent in Europe and parts of Asia \cite{RAMIREZ_NEMEurope:17EnergyPolicy}. This section presents an inclusive NEM analytical model for retail tariffs in the distribution systems \cite{Alahmed_Tong:22IEEETSG}.

\subsection{Revenue Metering}\label{subsec:MeteringPolicy}
The retail tariff applies to quantities measured by revenue meters. Modern smart meters provide bidirectional digital measurements of power flows on 5 to 60 minutes intervals for billing purposes \cite{DOE:16Rpt}. The power measurement interval defines the finest pricing resolution.

\begin{figure}[tb]
    \centering
    \includegraphics[scale=0.5]{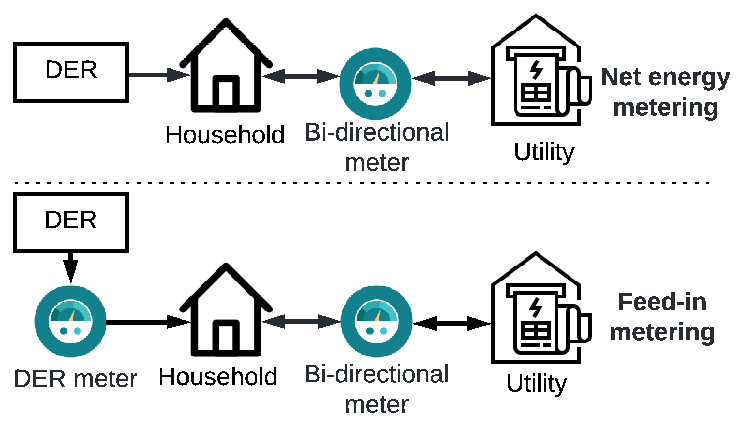}
    \caption{Net energy metering and feed-in metering.}
    \label{fig:FiTNEMcombineed}
\end{figure}

 The top panel of Fig.\ref{fig:FiTNEMcombineed} shows the revenue meter setting for the NEM tariff, which involves a bidirectional energy meter that measures the customer's {\em net energy consumption} within the meter's sampling interval. Under NEM, neither the gross consumption nor the BTM DER generation of a prosumer is observable to the utility\footnote{In some cases, the utility deploys sub-meters to enable measuring the BTM generation for performance-based incentives (PBI), utility planning,  and tracking climate goals.}.  In general, such load masking hinders the utility's understanding of customers' consumption patterns and BTM DER operational efficiency \cite{Brown_Reliability:17Book}.

Under FiT, the customer sells its gross DER generation to and buys its gross household consumption from the utility \cite{YamamotoPricingEF:12SE}. Therefore, the feed-in metering produces two registers for gross consumption and DER generation using two physical meters.  Shown in the bottom panel of Fig.\ref{fig:FiTNEMcombineed} is one of the possible configurations that enable the reconstruction of gross DER generation and household consumption. Some FiT schemes require the DER to be physically disconnected from the household loads \cite{Zinman_MeteringArchitectures:17NREL}, but the FiT payment model is formulated regardless of the physical connection variants.

\subsection{Retail tariff model}\label{subsec:utilityRate}
\subsubsection{Rate-setting, net-billing, and consumption decision periods}
\begin{figure}[tb]
    \centering
    \includegraphics[scale=0.45]{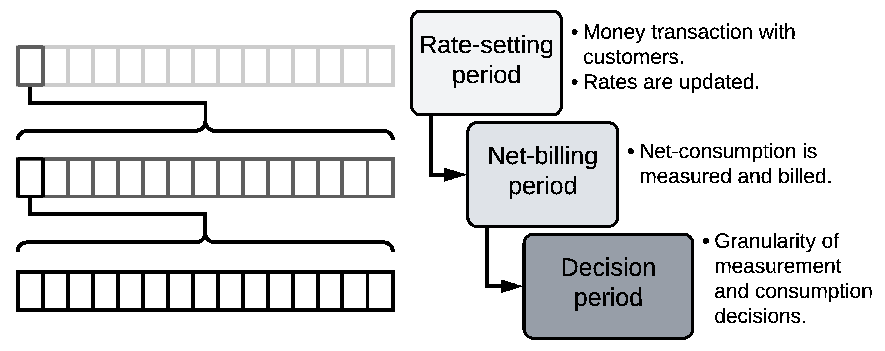}
    \caption{Rate-setting, net-billing, and customer decision periods.}
    \label{fig:TemporalTerm}
\end{figure}

 We define three decision epochs that affect the short and long-run analysis of retail tariff policies, illustrated in  Fig. \ref{fig:TemporalTerm}.  The  {\em rate-setting period} defines the frequency of rate updates by the regulator\footnote{The regulator is usually the Public Utilities Commission (PUC), and the rate-setting period can range from monthly, seasonal or annual depending on the PUC regulations.}.  We assume that, within a rate-setting period, the consumption and compensation rates for electricity are fixed and known to the customers.

 Hughes and Bell \cite{Hughes&Bell:06EP} provided a comprehensive taxonomy of a large number of variations of compensation mechanisms and terminologies circa 2006.  More nomenclatures have been introduced since then to delineate detailed aspects of billing mechanisms.  With little loss of generality for our discussion, we define {\em net-billing period} as the duration within which the customer's net consumption is computed and settled in monetary values\footnote{With digital smart meter measurements, the net-billing period is defined by the number of meter measurements used to compute payment.}.  In California, for example, the NEM net-billing period is 60 minutes for residential customers and 15 minutes for commercial customers \cite{CPUC_NEM3}.   The net-billing period can be instantaneous (using a single power measurement) as implemented in Arizona and Indiana \cite{dsireArizona,IndianaNEM:21}.  A shorter net-billing period narrows the window that the BTM generation can offset consumption, reducing the "kilowatt-hour banking effect" effect that allows later consumptions to be offset by earlier DER generation. Successor NEM policies are expected to have more granular net-billing periods \cite{CPUC_NEM3,IndianaNEM:21}. 

Customer consumption decisions are constrained by the sensing and actuation resolution of the energy management system.  The DER type ({\it e.g.,\/} rooftop PV), consumption patterns, and resolutions of  DER production and atmospheric measurements all influence the granularity of the consumption decision periods.

\subsubsection{The NEM X tariff model}\label{subsec:NEMXpolicies}
NEM X is an inclusive model for net-metering tariffs with differentiated  (customer) import and export prices \cite{Alahmed_Tong:22IEEETSG}.  It includes, as special cases, most NEM tariffs that have been implemented and those under consideration.  In its most generic form, the customer payment schedule $ P^{\mbox{\tiny NEM}}_\pi$  under NEM X in a single billing period is based on the net consumption $z:=d-r$ with consumption $d$ and BTM generation $r$ given by
\begin{equation}
\label{eq:netbasedpayment}
   P^{\mbox{\tiny NEM}}_\pi(z) =  \pi^{+}z \chi{(z)}+ {\pi}^{-}z (1-\chi{(z)}) +\pi^0,
\end{equation}
where $\chi(x)=1$ if $x\geq 0$ and $\chi(x)=0$ otherwise. The tariff parameter $\pi=(\pi^+,\pi^-,\pi^0)$ defines the retail rate $\pi^+$ when the prosumer is a \textit{net consumer} and the compensation rate (a.k.a. sell or export rate) $\pi^-$ when the prosumer is a \textit{net producer}. The volume-independent fixed charge $\pi^0$ represents a ``grid connection charge'', ranging from
 \$0 to over \$30 a month \cite{Borenstein&Bushnell_DoTwoElectricityPrices:18NBER}.

The first generation NEM policies (NEM 1.0) set $\pi^+=\pi^-$, which compensates a customer for its export at the retail rate for its import,  serving as a strong incentive for rooftop solar adoption.  An exception of the NEM X model is the decommissioned NEM 1.0 in parts of California, which used an inclining-block rate (IBR) rather than a linear (flat) rate.

The second generation of NEM tariff policies (NEM~2.0) is considered transitory as states adjust their tariff policies toward a more sustainable setting.  NEM 2.0 includes several variations, most significant being the differentiated retail ($\pi^+$) and compensation ($\pi^- < \pi^+$) rates.  The compensation rate $\pi^-$, also referred to as the marginal value of DER (VDER\footnote{Also referred to as the value-of-solar (VOS), when referring exclusively to solar PV.}),  quantifies the benefits gained or costs avoided due to the BTM DER \cite{Hayibo_VoS:21RSER,BLACKBURN_VOS:14EJ,CPUC_NEM3}. Note that, under NEM X, the compensation rate $\pi^-$ only applies to the portion of BTM generation exceeding the customer's consumption,  which implies that self-consumed DER production is virtually compensated at the retail rate $\pi^+$. Further export rate reductions are favorable by utilities to better align the cost of purchasing excess generation and the cost of securing that same amount from the wholesale electricity market. The cost difference is attributed to the fact that prosumers face a retail price that bundles multiple utility costs including capacity and fixed costs, which do not proportionally scale down when prosumers become exporters. This incurs a cost to the utility that is usually shifted to other customer types to achieve revenue adequacy or the approved profit margin.

NEM 3.0 represents the less well-defined future next generation of NEM policies, including lowered compensation rates and discriminative components based on the capacity of the BTM DERs and income levels.  See discussions in Sec.~\ref{sec:policyvariation}.

\subsubsection{FiT X models} FiT separately prices the gross consumption and generation.  With the NEM X notation,  the payment under FiT is given by
\begin{equation}
\label{BASApayment}
     P_{\pi}^{\mbox{\tiny FiT}}(d,r) =  \pi^+ d - \pi^-  r+\pi^0.
\end{equation}
It turns out that the NEM and FiT policies differ by
\begin{equation}
\label{eq:PaymentsDifference}
     P_{\pi}^{\mbox{\tiny FiT}}(d,r) - P_{\pi}^{\mbox{\tiny NEM}}(d-r) = (\pi^+-\pi^-)\min{\{d,r\}}.
\end{equation}
The two policies are identical when $\pi^+ = \pi^-$. When $\pi^+> \pi^-$, the customer payment under NEM is lower than that under FiT, making NEM X a stronger incentivizing policy for DER adoption.  On the other hand,  FiT policy can set the compensation rate higher than the retail rate as implemented in parts of Europe, providing stronger incentives for DER adoption than that under NEM, which led to the high solar penetration in Germany \cite{IRENA_NetBillingSchemes:19IRENA}.

\subsubsection{Policy variations}\label{sec:policyvariation}
We focus on the policy variations of the NEM X tariff model, which mostly apply to the FiT X tariff model. The baseline NEM X model defined in (\ref{eq:netbasedpayment}) can be generalized in multiple dimensions; some of these variations are being considered for the successors of existing NEM tariffs.  The first type of variation makes the NEM X parameter $\pi$ time-varying.  Under the time-of-use (TOU) model, separate rate parameters are defined for the peak and off-peak periods.  NEM X model under TOU has been implemented in California \cite{cpuc_NEM2}, Arizona \cite{NEMstateReview:20Dominion}, and Nevada \cite{NEM_Nevada}.  Beyond the TOU framework, NEM X can be easily generalized for the real-time (dynamic) retail pricing \cite{Borenstein:05EJ}. 

Another class of variations makes NEM discriminative across multiple customer groups. For instance, the fixed charge  $\pi^0$ can be discriminatory by making it dependent on income \cite{Next10Report}, DER capacity \cite{CBC_NYstate, CPUC_NEM3}, or customer sub-class of
consumer and prosumer groups.  In California, a special grid-access charge (GAC) has been proposed for prosumers as part of NEM 3.0 \cite{CPUC_NEM3}.  In New York, a DER capacity-dependent fixed charge is proposed \cite{CBC_NYstate}.  Moreover, income-based fixed charges have been proposed to address equity issues \cite{Next10Report}.  Additionally, a potential step in NEM 3.0 is to treat prosumers and consumers as two distinct customer classes with different retail rates \cite{Trabish_volumetricdiscrimination:17UtilityDive,StantonNEMreview:19NARUC}.

\subsubsection{Related tariff models terminologies}\label{subsec:retailTariff}

The NEM X and FiT tariff models defined in (\ref{eq:netbasedpayment}--\ref{BASApayment}) cover most existing NEM and FiT implementations that often have different terminologies. . 
Earlier literature uses {\em full NEM} for equal retail and export rates ($\pi^+=\pi^-$) and {\em partial NEM} when $\pi^-<\pi^+$ \cite{Darghouth_NEMFeedbackloops:16AE}. When the net-billing period is long (e.g. monthly, bi-annually, or annually), the NEM X is approximately NEM 1.0 \cite{YamamotoPricingEF:12SE, Tong_DERdynamics:20TAC,BorensteinPrivateNetBenefits:15NBER}, where the customer, by the end of a billing period, is very likely a net-consumer \cite{Wolfram_BillingPeriod:16Haas,Boero_DERmicroeconomics:16SE}. When the net-billing period is an integer multiple of power measurement period, NEM X is the so-called \textit{net-billing} tariff \cite{Zinman_MeteringArchitectures:17NREL,CPUC_NEM3,Darghouth_PVSolar:19LBNL,lawsonNEM:19Congress,Hughes&Bell:06EP}, and \textit{net-FiT} \cite{NETFiT:18CleanEnergySolutions,Zinman_MeteringArchitectures:17NREL}.  When the net-billing period equals the measurement period, NEM X becomes the discrete-time version of the \textit{net-purchase and net-sale} policy \cite{YamamotoPricingEF:12SE,Varaiya_NEMA:19TSG}. The FiT X model in (\ref{BASApayment}) is sometimes called \textit{buy-all sell-all} tariff \cite{Zinman_MeteringArchitectures:17NREL}, \textit{gross-FiT} \cite{GrossFit:18AustralianEnergyCouncil}, and \textit{value of solar tariff (VOST)} \cite{Kirschen_VOST:15PESGM,Wolfram_BillingPeriod:16Haas}.
\par A summary of the presented two tariff models is provided in Table~\ref{tab:NetFitcomparison}.

\begin{table}[tb]
\centering
\caption{Summary of retail tariff models.}
\label{tab:NetFitcomparison}
\resizebox{0.49\textwidth}{!}{%
\begin{tabular}{@{}ccc@{}}
\toprule \toprule
Tariff                & NEM X                                                                                                                 & FiT X                                                                                                                  \\ \midrule
Tariff model          & $P^{\mbox{\tiny NEM}}_\pi(z) =  \pi^{+}z \chi{(z)}+ {\pi}^{-}z (1-\chi{(z)}) +\pi^0$                                  & $P_{\pi}^{\mbox{\tiny FiT}}(d,r) =  \pi^+ d - \pi^-  r+\pi^0$                                                          \\
Other terminology & \begin{tabular}[c]{@{}c@{}}NEM 1.0, NEM 2.0, full NEM, partial NEM,\\ net-billing, net-FiT\end{tabular}               & Buy-all, sell-all, gross FiT                                                                                           \\
Self-consumption      & Yes                                                                                                                   & No                                                                                                                     \\
Meters needed         & 1                                                                                                                     & 2                                                                                                                      \\
Pros                  & \begin{tabular}[c]{@{}c@{}}Implementation simplicity and low cost, \\ self-consumption, back-up services\end{tabular} & \begin{tabular}[c]{@{}c@{}}Separate consumption and generation pricing, \\ measurable DG performance\end{tabular}      \\
Cons                  & Load masking, grid defection, cost-shifts                                                                             & \begin{tabular}[c]{@{}c@{}}Higher reverse power flows, \\ illegal self-wiring, higher implementation cost\end{tabular} \\ \bottomrule \bottomrule
\end{tabular}%
}
\end{table}

\section{Optimal prosumer decisions}
\label{sec:prosumer}
We present in this section the structure of the optimal prosumer decision under the NEM X and FiT policies and the near closed-form characterization of the optimal consumption.  The results shown here are built upon the work of \cite{Alahmed_Tong:22IEEETSG} with new considerations (in Sec.~\ref{subsec:DERuncertainty}) when the BTM generation is stochastic.

\subsection{Prosumer decision under NEM X} \label{subsec:prosumerNEM}
Consider a prosumer's energy management system involving $M$ devices facing NEM X tariff with parameter $\pi=(\pi^+,\pi^-,\pi^0)$.  Let ${\bm d}=(d_1,\cdots, d_M)$ be the consumption bundle of $M$ devices,  $U({\bm d})$ the utility of consumption, and $r$ the BTM generation. We call a prosumer {\em active} if its consumption decision is a function of $r$ and {\em passive} otherwise.

A surplus-maximizing active prosumer solves the following optimization
\begin{equation}
 \begin{array}{lll}
\mathcal{P}_{\mbox{\tiny NEM X}}: &  \underset{\bm{d}  \in \mathbb{R}^M}{\rm maximize}& J_\pi^{\mbox{\tiny NEM}}(\bm{d}) :=U(\bm{d}) - P_{\pi}^{\mbox{\tiny NEM}}(z) \\
&{\rm subject~ to} & z = \sum_{i=1}^M d_i -r \\
 & & \bm{0} \preceq   \bm{d}  \preceq \bar{\bm{d}},
 \end{array}
 \label{eq:prosumerOPT_NEM}
 \end{equation}
 where $\bar{\bm{d}}$ is the consumption upper limit\footnote{Typically, the household electricity  consumption is reasonably small, and the household budget constraint is ignored in our formulation.}.

For a concave utility\footnote{Without loss of generality and to gain a cleaner structure, the theoretical results assume strict concavity of $U(\cdot)$.} function $U(\cdot)$, the above optimization is convex, though non-differentiable.  It turns out, however,  that the optimal consumption bundle has a near closed-form solution as given by Theorem~\ref{thm:structure} in Appendix \ref{sec:appendixProofs}.  Here we describe the general structure of the optimal consumption bundle $\bm{d}^\ast$ as a function of the BTM DER level $r$ and the intuitions behind the threshold-based policy.

\subsubsection{Structures of optimal consumption}  Under NEM X,  the optimal consumption decision is a  {\em two-threshold policy} with thresholds $(d^+,d^-)$  computed  {\it a priori} from the NEM X parameters and the {\em marginal utility function} $V_i(x) := \frac{d}{dx_i}U(x)$ by
\begin{align}
d^+ &:= \sum_i  \max \{0,\min\{V_i^{-1}(\pi^+),\bar{d}_i\}\},\\
d^- &:= \sum_i \max  \{0,\min\{V_i^{-1}(\pi^-),\bar{d}_i\}\} \ge d^+.
\end{align}
Note that $d^+,d^-$ are independent of DER $r$ and uniform across all devices.

\begin{figure}[tb]
    \centering
    \includegraphics[trim=0cm 3cm 0cm 2cm, scale=0.25]{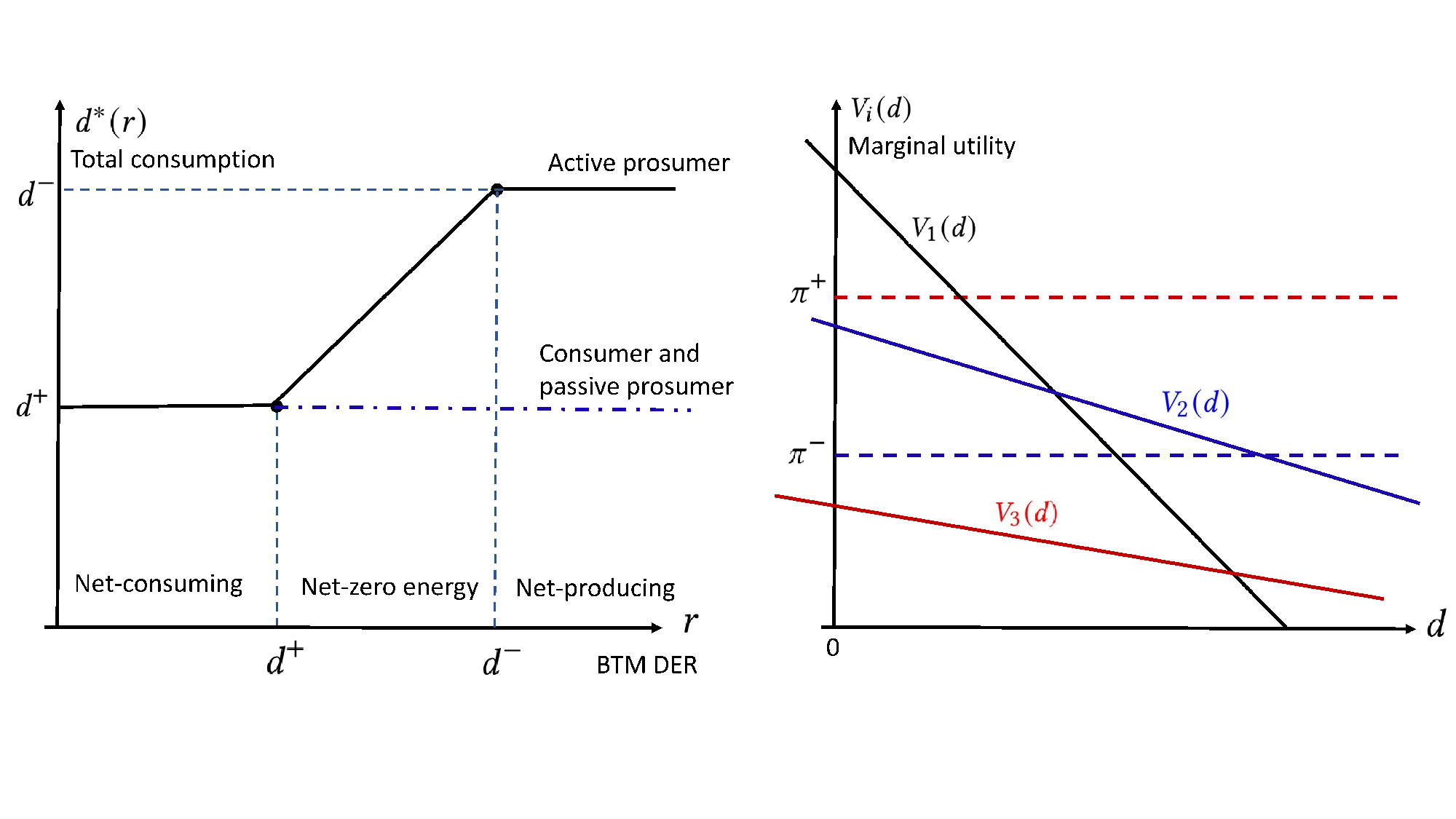}
    \caption{Left: optimal prosumer consumption and consumption zones under NEM X.  Right: Consumption allocation to devices based on marginal utilities. }
    \label{fig:zone}
\end{figure}

With thresholds $(d^+, d^-)$, the optimal prosumer consumption policy is to partition the decision space on DER into three operation zones   illustrated on the left panel of Fig.~\ref{fig:zone} and defined by
\begin{enumerate}
\item  The {\it net-consumption zone} when $r < d^+$,  where the prosumer is a {\em net consumer;}.
\item  The {\it net-production zone} when $r>d^-$, where the prosumer is a {\em net producer;}.
 \item The {\it net-zero zone}  when $d^+\le r\le d^-$, where the prosumer consumes at the level of  BTM DER as a net-zero consumer.
\end{enumerate}

%

The optimal consumption allocation to each device depends on its marginal utility $V_i$ of consumption.  Specifically, the optimal consumption at device $i$ is given by
\begin{equation} \label{eq:d*}
d_i^*(r) = \left\{\begin{array}{ll}
d_i^+ := \max \{0,\min\{V_i^{-1}(\pi^+),\bar{d}_i\}\}, & r < d^+\\
d^-_i := \max  \{0,\min\{V_i^{-1}(\pi^-),\bar{d}_i\}\},  & r > d^-\\
d^o_i : = \max\{0, \min\{V_i^{-1}(\mu^*(r)),\bar{d}_i\}\}, & \mbox{o.w.},
\end{array}\right.
\end{equation}
where $d_i^+ \le d_i^o \le d^-_i$, and $\mu^*(r) \in [\pi^-, \pi^+]$ is a solution of $$\sum_{i=1}^M \max\{0, \min\{V_i^{-1}(\mu),\bar{d}_i\}\} = r.$$

Because FiT prices gross DER generation separately from gross consumption, a prosumer's consumption under FiT is independent of $r$, and is given by the solution above with $r=0$.

The implications of the optimal consumption structure are significant.  First, the thresholds are global to all devices and easily set (possibly in hardware).
The optimal consumption level can be implemented easily with minor complications in the net-zero zone, which gives a significant advantage in scalability over solving $\mathcal{P}_{\mbox{\tiny NEM X}}$ directly.

Second,  the near closed-form characterization of the optimal prosumer decision allows us to investigate how NEM parameters affect consumption behavior.  Furthermore,  while the existence of net-consumption and net-production zones are quite natural,  the existence of a positive-sized net-zero zone where prosumers are ``off the grid'' is particularly intriguing. The intuition is that, because $\pi^+ \ge \pi^-$, DER generation is more valuable consumed for greater utility than exported to the grid. The strategy for a surplus-maximizing prosumer is to minimize the amount of DER generation exported to the grid until the concavity of the utility limits the return of self-consumption.  A comparative static analysis (Theorem 2 in \cite{Alahmed_Tong:22IEEETSG}) shows that as the compensation rate $\pi^-$ decreases, the net-zero zone expands, creating a wider region that a group of prosumers neither generating nor consuming, which alleviates potential network congestion.

Finally, the optimal consumption allocation given by (\ref{eq:d*}) suggests that
devices with high marginal utility are scheduled to consume more, and devices with marginal utilities below the threshold set by $\pi$ are not scheduled.  As illustrated in the right panel of Fig.~\ref{fig:zone}, in the net-consumption zone when BTM DER is limited, only devices with marginal utilities at zero greater than the retail rate ($V_i(0)>\pi^+$) are scheduled (device 1 in the figure).  In the net-production zone when BTM DER is plenty, all devices with marginal utilities at zero greater than the compensation rate ($V_i(0)>\pi^-$) are scheduled (devices 1 and 2).  Those devices with marginal utilities at zero below the compensation rate ($V_i(0) \leq \pi^-$) are not scheduled (device 3).

\subsubsection{Intuition of optimal consumption structure}
The characterized threshold-based optimal consumption policy has an intuitive proof. Here, highlights on the intuition and insights are provided. 
\par Given the net-consumption indicator function ($\chi(z)$) in (\ref{eq:netbasedpayment}), the prosumer is either importing from the grid ($\chi(z)=1$), or exporting to the grid ($\chi(z)=0$). The rationale behind the three operation zones can be acquired from the following two DER-independent optimizations (assuming $M=1$):
\begin{align}
    d^{+} &:=\arg \max_{d}\; J_{r}^{+}(d):= U(d)-\pi^+ (d-r)\\
    d^{-} &:=\arg \max_{d}\;  J_{r}^{-}(d):= U(d)-\pi^- (d-r),
\end{align}
which, given the monotonicity of the marginal utility function $V(\cdot)$ and $\pi^+\geq \pi^-$, yields $d^-\geq d^+$. Note that $d^+$ and $d^-$ are independent of the BTM renewable $r$, which implies the prosumer consumption level is independent of $r$ as long as it is either net-producing or net-consuming. 

Consider the optimal consumption decision of the prosumer as $r$---the available BTM generation---increases.  Because $d^+$ is the optimal consumption level when the prosumer net-consumes, the prosumer achieves increasingly higher surplus as its payment is reduced by $r$, approaching to the highest surplus of $U(d^+)$ as $r\rightarrow d^+$.  

When $r$ grows slightly greater than $d^+$, because the utility function is monotonically increasing, the prosumer benefits from increased utility with higher consumption without increased payment as long as its increased consumption ($d^+$) matches with $r$.  However, increasing consumption while matching with $r$ cannot continue indefinitely because the utility is concave and its marginal utility decreases with $d$. When the benefit of increased consumption is below the benefit of exporting $r$ defined by $\pi^-$ (which first happens at $r=d^-$), the optimal consumption is fixed at $d^-$ thereafter and all excess BTM generation is exported to the grid.

\subsubsection{Prosumer surplus characteristics}
\begin{figure}[tb]
    \centering
    \includegraphics[scale =0.3]{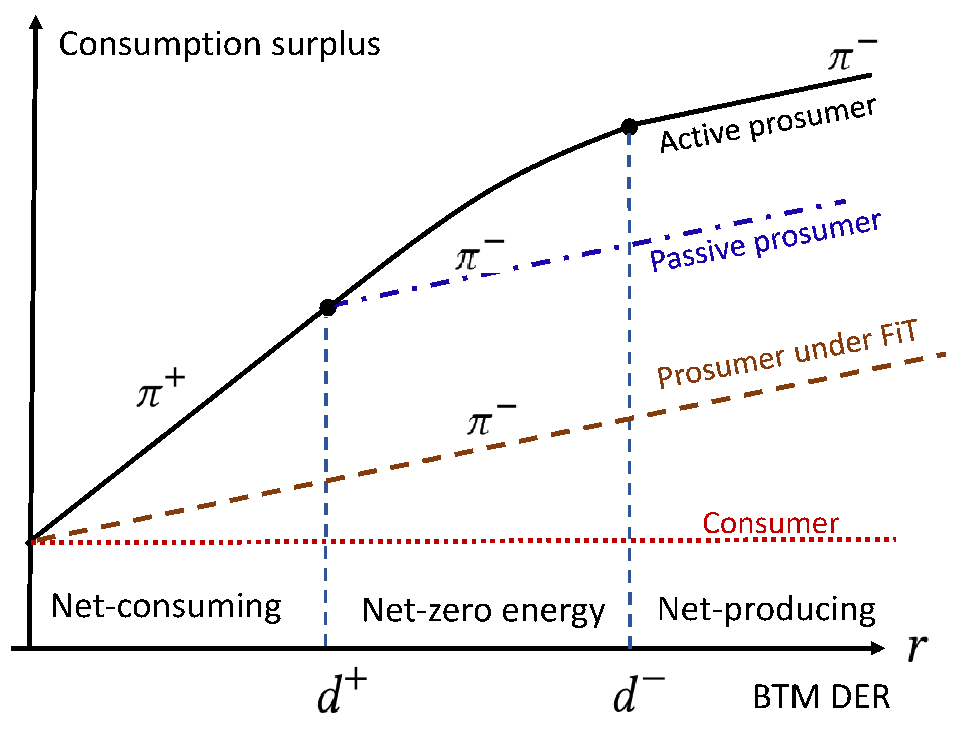}
    \caption{Customer surpluses.  Dotted (red)--Consumer surplus.  Dashed (brown)--Prosumer surplus under FiT.  Dash-dotted (blue): Passive prosumer under NEM.  Solid (black): Active prosumer under NEM. }
    \label{fig:surplus}
\end{figure}

The near closed-form solution (\ref{eq:d*}) of the prosumer optimal consumption makes it possible to characterize and compare surpluses under different decision models as illustrated in Fig.~\ref{fig:surplus}.

Consumers achieve the lowest surplus (dotted line) without DER, which corresponds to the case when $r=0$.   The passive prosumers' consumption is not a function of the available DER $r$.  Its surplus grows with $r$ at the rate of $\pi^+$ in the net-producing zone as DER is equivalently priced at $\pi^+$.  Beyond the net consumption zone, its surplus grows linearly with $r$ at the rate of $\pi^-$.
The active prosumer achieves the highest surplus that grows at the rate of $\pi^+$ in the net-consumption zone and $\pi^-$ in the net-production zone.  In the net-zero energy zone, the active total prosumer's consumption matches that of the renewable, and the total surplus grows in a nonlinear fashion until $r=d^-$.
The surplus of a prosumer under FiT grows linearly with the compensation rate $\pi^-$.

\subsection{Prosumer consumption under DER uncertainty}\label{subsec:DERuncertainty}
The optimal prosumer consumption decision in (\ref{eq:prosumerOPT_NEM}) assumes accurate measurement of the available BTM DER $r$ to allocate consumptions in each device over the net-billing period.  When the net-billing period is relatively long, the real-time implementation of the optimal decision at time $t$ requires a forecast of future DER output that can be inaccurate.  The optimal consumption under DER uncertainty becomes one of stochastic dynamic programming facing the classic ``curse of dimensionality''.
With the near closed-form solution in (\ref{eq:d*}), it is natural to consider a suboptimal but effective approach of model-predictive-control (MPC) strategies.

Consider the problem of scheduling $M$ devices within one net-billing period.  Assume that within one net-billing period, there are  $T$  sensing and control intervals.  Let the BTM DER within the net-billing period be $r_t, t=1, \ldots, T$.  At time $t=k$, the EMS has the realized DER  outputs $r_1,\ldots, r_{k-1}$ and forecasted DER outputs $\hat{r}_{k}, \ldots, \hat{r}_T $, and it has already exercised its consumption decisions up to time $k-1$.

 The MPC strategy calls to determine the consumption for the rest of the net-billing period, based on the realized and forecasted DER, and implements the actual allocation at time $t=k$.  Let the total realized DER, exercised consumption, and utility  up to time $k$ be $\tilde{r}_k$,  $\tilde{d}_k$, and $\tilde{U}_k$, respectively,  given by
 \[
 \tilde{r}_k := \sum_{t=1}^{k-1} r_{t},~\tilde{d}_k:= \sum_{i=1}^M \sum_{t=1}^{k-1} d^\pi_{t,i},~
\tilde{U}_k :=\sum_{i=1}^M \sum_{t=1}^{k-1} U_{t,i}(d^\pi_{t,i}).
\]
The MPC optimization at $t=k$ is given by
 \begin{equation}
 \begin{array}{lll}
\mathcal{P}^{\mbox{\tiny MPC}}_{\mbox{\tiny NEM X}}: &  \underset{\{d_{t,i}, \forall t\ge k, i\}}{\rm minimize}
&P_{\pi}^{\mbox{\tiny NEM}}\big(\sum_{t=k}^T (\sum_{i=1}^M d_{t,i}-\hat{r}_t) + \tilde{d}_k \\
  && -\tilde{r}_k\big) -\sum_{i=1}^M\sum_{t=k}^T U_{t,i}(d_{t,i})\\
 & \mbox{subject to} & \bar{d}_{t,i} \ge d_{t,i} \ge 0,~~\forall t\geq k, \forall i.
 \end{array}
 \label{eq:prosumerOPT1}
 \end{equation}
Leveraging the solution of the one-shot optimization (\ref{eq:d*}), the solution of $\mathcal{P}^{\mbox{\tiny MPC}}_{\mbox{\tiny NEM X}}$ can also be obtained in near closed-form.   See Theorem~\ref{thm:MPCversionofOCT} and a prototype implementation in Appendix~\ref{sec:appendixProofs}.

\section{Regulator's rate-setting Decision}
\label{sec:RegulatorProblem}
This section considers the regulator's rate-setting decision process based on a system theoretic model shown in Fig. \ref{fig:Dynamics}, which characterizes the endogenous interaction of tariff parameters and DER adoption rates, following the general approach in \cite{Tong_DERdynamics:20TAC}.

\subsection{A feedback model for long-run DER adoption}
\begin{figure}[tb]
    \centering
    \includegraphics[scale=0.45]{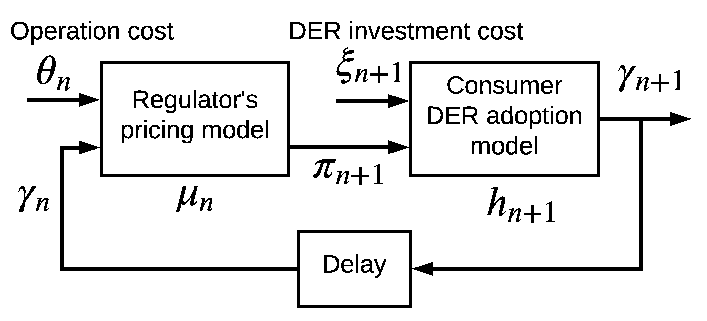}
    \caption{Nonlinear feedback model of DER adoption.}
    \label{fig:Dynamics}
\end{figure}

We present a feedback system model that captures the long-run dynamics of the rate-setting process illustrated in Fig.~\ref{fig:Dynamics}.

Let the sequence of rate-setting periods be indexed by $n$.  In the $n$th period that determines the retail tariff parameters in period $n+1$, the utility presents to the regulator the expected fixed cost $\theta_{n}$ in period $n+1$ and the current fraction of BTM DER adoption $\gamma_n\in [0,1]$ (with $\gamma=0$ for no adoption).  The regulator sets the NEM X  rate parameter $\pi_{n+1}$ based on a certain social welfare criterion subject to that the regulated utility  recovers its costs $\theta_{n}$:
\begin{equation}\label{eq:NextPeriodTariff}
    \pi_{n+1} = \mu_n(\gamma_n,\theta_{n}).
\end{equation}
With the price set at $\pi_{n+1}$ for the period $n+1$, consumers (without DER) decide whether to adopt BTM DER based on $\pi_{n+1}$ and the cost of investment $\xi_{n+1}$, resulting in a new level of adoption $\gamma_{n+1}$  at the end of period $n+1$ according
to a certain adoption model $h_{n+1}$:
\[
\gamma_{n+1} = h_{n+1}(\pi_{n+1},\xi_{n+1}).
\]
The regulator's rate-setting period and consumer's DER adoption models are discussed next.

\subsection{The short-run rate-setting model}
We assume that the regulator follows the principle of Boiteux-Ramsey pricing to set NEM X parameter $\pi$, acknowledging that, in practice, the regulator incorporates many factors in the rate-setting process \cite{Brown&Sibley:1986book}. In particular, the Boiteux-Ramsey pricing maximizes the overall social welfare subject to that the utility recovers its cost over the rate-setting period  \cite{Boiteux:56EC,RamseyPricing:1927EJ}.

Using a representative prosumer with optimal consumption decisions (\ref{eq:d*}), a stochastic Boiteux-Ramsey pricing optimization involving $J$ net-billing periods can be formulated as in \cite{Alahmed_Tong:22IEEETSG}:
\begin{equation}
 \begin{array}{ll}
\mu^\ast_n :\underset{\boldsymbol{\pi} }{\rm maximize}& \sum_{j=1}^J \mathbb{E}\bigg(S_{c}^\pi (r_j,\gamma_n)+ \gamma_n \mathcal{E}(r_j)\bigg)\\
\mbox{subject to} & \sum_{j=1}^J \mathbb{E}(S_{u}^\pi (r_j,\gamma_n,\theta_n)) =0,
\label{eq:Ramseymaximization}
 \end{array}
\end{equation}
where the expectation is taken over random DER generations over $J$  {\em net-billing periods}\footnote{Note that $r_j$ in the $j$th interval is different from $r_k$ in Sec.~\ref{subsec:DERuncertainty} for the $k$th measurement interval.}.    In (\ref{eq:Ramseymaximization}),
 $\mathcal{E}$ is the environmental benefits brought by BTM DER. The customer and utility company surpluses, $S_c^\pi$ and $S_u^\pi$, are defined next.

The customer surplus in the $j$th net-billing period is the sum of adopters and non-adopters surpluses given by
\begin{equation}
S_{c}^\pi (r_j,\gamma_n)) = \gamma_n S^\pi(r_j) + (1-\gamma_n)S^\pi(0),
\label{eq:Sc}
\end{equation}
where the first term on the right-hand side $S^\pi(r) :=J_\pi^{\mbox{\tiny NEM}}({\bm d}^*(r))$ is the maximum prosumer surplus from the prosumer surplus-maximization (\ref{eq:prosumerOPT_NEM}), and the second term is the consumer surplus by setting the BTM DER generation to zero.

The utility's surplus in the net-billing period $j$ is given by the income from its customers minus the variable energy cost from the wholesale electricity market and  the anticipated fixed operating cost $\theta_{n}$ in period $n+1$:
\begin{align}
\label{eq:Su}
S_{u}^\pi (r_j,\gamma_n)) &=   \gamma_n P_{\pi}^{\mbox{\tiny NEM}}(d^*(r_j)-r_j) + (1-\gamma_n) P_{\pi}^{\mbox{\tiny NEM}}(d^*(0))\nonumber\\
   & - C(y_j(r_j,\gamma_n)))-\theta_n/J,
\end{align}
where  $d^*(r_j)=\sum_i d_i^*(r_j)$ is the total consumption in the net-billing period $j$ and
$C$   the utility’s variable cost function to
meet the total customers’ net demand $y_j$ defined by
\[
y_j(r_j,\gamma_n) := \gamma_n (d^*(r_j)-r_j)+ (1-\gamma_n)d^*(0).
\]

\subsection{DER adoption model}\label{subsec:AdoptionModel}
As in \cite{Tong_DERdynamics:20TAC}, we adopt the widely used $S$-curve technology diffusion model for the consumer DER adoption \cite{bassModel:1969MS}, where various factors influecing adoption decisions are embedded in the shape of the $S$-curve. The \textit{DER adoption curve} $s(t;\pi,\xi)$ is a function of  the rate-setting period index $t$, parameterized by the retail tariff parameter $\pi$ and the cost of DER adoption $\xi$:
\begin{equation}\label{eq:AdoptionCurve}
    s(t;\pi,\xi) = \eta_\infty(\pi,\xi) \eta(t),
\end{equation}
where $\eta_\infty(\pi,\xi)$ is the DER \textit{market potential}, and $\eta(t)$ is a sigmoid function that models the \textit{cumulative installed fraction} satisfying $\eta(0)=0$ and $\lim_{t \rightarrow \infty} \eta(t)=1$.

\par The Bass diffusion model \cite{bassModel:1969MS} is therefore a special case of (\ref{eq:AdoptionCurve}). The adoption curve given by (\ref{eq:AdoptionCurve}) defines adoption evolution when tariff parameter $\pi$ is set exogenously and DER cost $\xi$ are fixed.  To capture the diffusion dynamics, the retail tariff parameters need to be set endogenously by the feedback dynamic model in Fig.~\ref{fig:Dynamics}
\begin{align}\label{eq:dynamicallEquationforGamma}
    h_{n+1}&: \gamma_{n+1}=\nn\\& \begin{cases}\gamma_{n},   \text { if } \eta_{\infty}\left(\pi_{n+1}, \xi_{n+1}\right)<\gamma_{n} ; \\ s\left(1+\eta^{-1}\left(\frac{\gamma_{n}}{\eta_{\infty}\left(\pi_{n+1},\xi_{n+1}\right)}\right) ; \pi_{n+1}, \xi_{n+1}\right),  \text {o.w., }\end{cases}
\end{align}
which, together with (\ref{eq:NextPeriodTariff}), specifies a nonlinear dynamic model. The stability of this model is analyzed for NEM 1.0 in \cite{Tong_DERdynamics:20TAC}.

\section{Social welfare, cross subsidies, and Market potential}\label{sec:adoptionModelDynamics}
We evaluate the performance of NEM tariffs under three related but sometimes conflicting objectives:  efficiency (social welfare), cross-subsidy (cost-shifts), and the rate of DER adoption.

\subsection{Social welfare}
We adopt a generalized notion of the expected social welfare as in the rate-setting optimization (\ref{eq:Ramseymaximization}) that includes customer surplus, the utility surplus, and the externality of environmental benefits brought by the adoption of BTM DER:
\begin{equation}\label{eq:welfare}
W^{\pi}_{\gamma_n} := \sum_{j=1}^J \mathbb{E} \bigg( S_{c}^\pi (\bm{d}^*(r_j),\gamma_n) +S_{u}^\pi (r_j,\gamma_n,\theta_n)+ \gamma_n \mathcal{E}(r_j)\bigg),
\end{equation}
where the expectation is taken over stochastic DER generation. In (\ref{eq:welfare}), $\mathcal{E}(r_j) = \pi^e \mathbb{E}(r_j)$ is the environmental benefits of DER production with $\pi^e$ as its shadow price \cite{Wiser_EnvironmentalBenefits:16EnergyJournal}. Given the revenue adequacy constraint in (\ref{eq:Ramseymaximization}), the second term of the right hand side of (\ref{eq:welfare}) is zero.

From Fig.~\ref{fig:surplus}, the breakeven condition in (\ref{eq:Ramseymaximization}) and the linear form of environmental benefits, it is immediate that the short-run social welfare under fixed $\pi$ increases with $r$. For a long-run analysis, however, the NEM tariff $\pi$ is endogenously determined as a function of the DER adoption. The social welfare may very well decrease with $r$ because the break-even condition of the Ramsey pricing makes it necessary to increase the retail price $\pi^+$, more prominently reducing the consumer surplus than the increase of prosumer surplus and environmental benefits. See the numerical results in Fig.~\ref{fig:NEMFiTNum} and discussions in Sec.~\ref{sec:numerical}.

\subsection{Cross-subsidy}  Through BTM DER, prosumers avoid a part of the payment that supports the overall grid operation, resulting in cross-subsidies\footnote{A game theoretic test of cross-subsidy is formulated by Faulhaber \cite{Faulhaber:75}, which implies that, in the absence of cross-subsidy,  the consumer group should not pay more than when it is served in absence of the prosumer.}  of prosumers by consumers.  A practical measure of cross-subsidies is the expected cost-shifts $\psi^\pi_{\gamma_n}$ from adopters to non-adopters defined by
\begin{equation}
    \label{eq:ProfitDefficit}
    \psi^\pi_{\gamma_n} = \sum_j \gamma_n  \mathbb{E}\left(\Delta P_{\pi}^{\mbox{\tiny NEM}}\left(r_j\right)- \pi^{\mbox{\tiny SMC}} r_j \right),
\end{equation}
where $\pi^{\mbox{\tiny SMC}}$ is the social marginal-cost pricing of electricity \cite{Next10Report}, and $\Delta P^{\mbox{\tiny NEM}}_\pi\left(r_j\right)$ is the bill savings due to onsite DER production given by
\begin{equation} \label{eq:billsaving}
    \Delta P_\pi^{\mbox{\tiny NEM}}(r_j):= P_\pi^{\mbox{\tiny NEM}}(d^*(0)) - P^{\mbox{\tiny NEM}}_\pi(d^*(r_j) - r_j),
\end{equation}
which is the difference between the payment under optimal consumption before and after installing the DER.
Cost-shifts occur when the bill savings of the adopters $\Delta P_\pi\left(r_j\right)$ exceeds the utility's avoided cost due to BTM DER generation \cite{EdisonReport_Borlick_Wood}.


\subsection{Payback time and market potential}
A major factor influencing consumer DER adoption is the investment {\em payback time} $T_{\mbox{\tiny PB}}^{\pi}(\xi)$, which depends on the cost of DER $\xi$ and retail tariff parameter $\pi$ that affects bill savings.  An estimate of the payback time assuming that the current tariff persists indefinitely is based on the expected bill-saving  (\ref{eq:billsaving}) as
\begin{equation}\label{eq:TNPpaybackTime}
    T_{\mbox{\tiny PB}}^\pi(\xi) =\min_{t^\ast}\left\{t^\ast: \sum_{t=0}^{t^\ast} \left(\frac{1-\nu}{1+\zeta}\right)^t  \mathbb{E}\left(\Delta P_{\pi,t}^{\mbox{\tiny NEM}}(r_t)\right) \geq \xi \right\},
\end{equation}
where $\nu,\zeta \in [0,1)$ are the BTM DER system degradation factor and interest rate, respectively.

Given $ T_{\mbox{\tiny PB}}^{\pi_n}(\xi_n)$ in rate-setting period $n$, the market potential in the  DER adoption curve (\ref{eq:AdoptionCurve})
is given by \cite{beckMktPotential:2009}
\begin{equation}\label{eq:MktPoten}
    \eta_\infty(\pi_n,\xi_n) = \alpha \exp\bigg({\mu T_{\mbox{\tiny PB}}^{\pi_n}(\xi_n)}\bigg),
\end{equation}
where $\alpha$ is the market size and $\mu$ the sensitivity of the payback time. 

\section{Numerical Results}
\label{sec:numerical}
We assume a hypothetical regulated utility company serving residential customers consisting of $\gamma$ fraction of (active) prosumers and $1-\gamma$ fraction of consumers.  We analyze the short and long-run adoption, cost-shifts, and welfare of rooftop solar adoption.  The detailed settings of the numerical evaluations, including household loads, rooftop solar, retail prices, utility company's fixed costs, and utility function parameters, are given in Appendix \ref{sec:appendixNumData}.

\subsection{Short-run performance analysis}
A short-run performance comparison between NEM X and FiT X policies was implemented under three different tariff parameter settings, where the same tariff parameter $\pi$ was applied to both NEM X and FiT X. The NEM 1.0 policy had $\pi^-=\pi^+$ and a flat rate.  The NEM 2.0 policy was similar to the Californian version, where a TOU rate\footnote{Similar to PG\&E TOU-B, the peak ratio is 1.5 and the peak period is 16 -- 21.} was used with a small price-differential $\pi^-=\pi^+-\$0.035$/kWh \cite{cpuc_NEM2}.  The NEM-SMC policy was an extension to 2.0, but with a value of solar exports equivalent to the SMC rate ($\pi^- = \pi^{\mbox{\tiny SMC}}$), as discussed in \cite{Borenstein&Bushnell_DoTwoElectricityPrices:18NBER}. In all policies, there are no fixed charges ($\pi^0=0$). The studied policies are summarized in Table \ref{tab:NumPolicySummary}.

\begin{table}[htbp]
\centering
\caption{Description of studied policies.}
\label{tab:NumPolicySummary}
\resizebox{0.47\textwidth}{!}{%
\begin{tabular}{@{}cccc@{}}
\toprule \toprule
\begin{tabular}[c]{@{}c@{}}NEM X\\ FiT X\end{tabular} & Rate & Export Rate                    & Notes                                  \\ \midrule
1.0                                                   & Flat & $\pi^-=\pi^+$                  &                                        \\
2.0                                                   & TOU  & $\pi^-=\pi^+-0.035$\$/kWh      & 1.5 peak ratio and 16 – 21 peak period \\
SMC                                                   & TOU  & $\pi^-=\pi^{\mbox{\tiny SMC}}$ & 1.5 peak ratio and 16 – 21 peak period \\ \bottomrule \bottomrule
\end{tabular}%
}
\end{table}

\begin{figure}[htbp]
    \centering
    \includegraphics[scale=0.325]{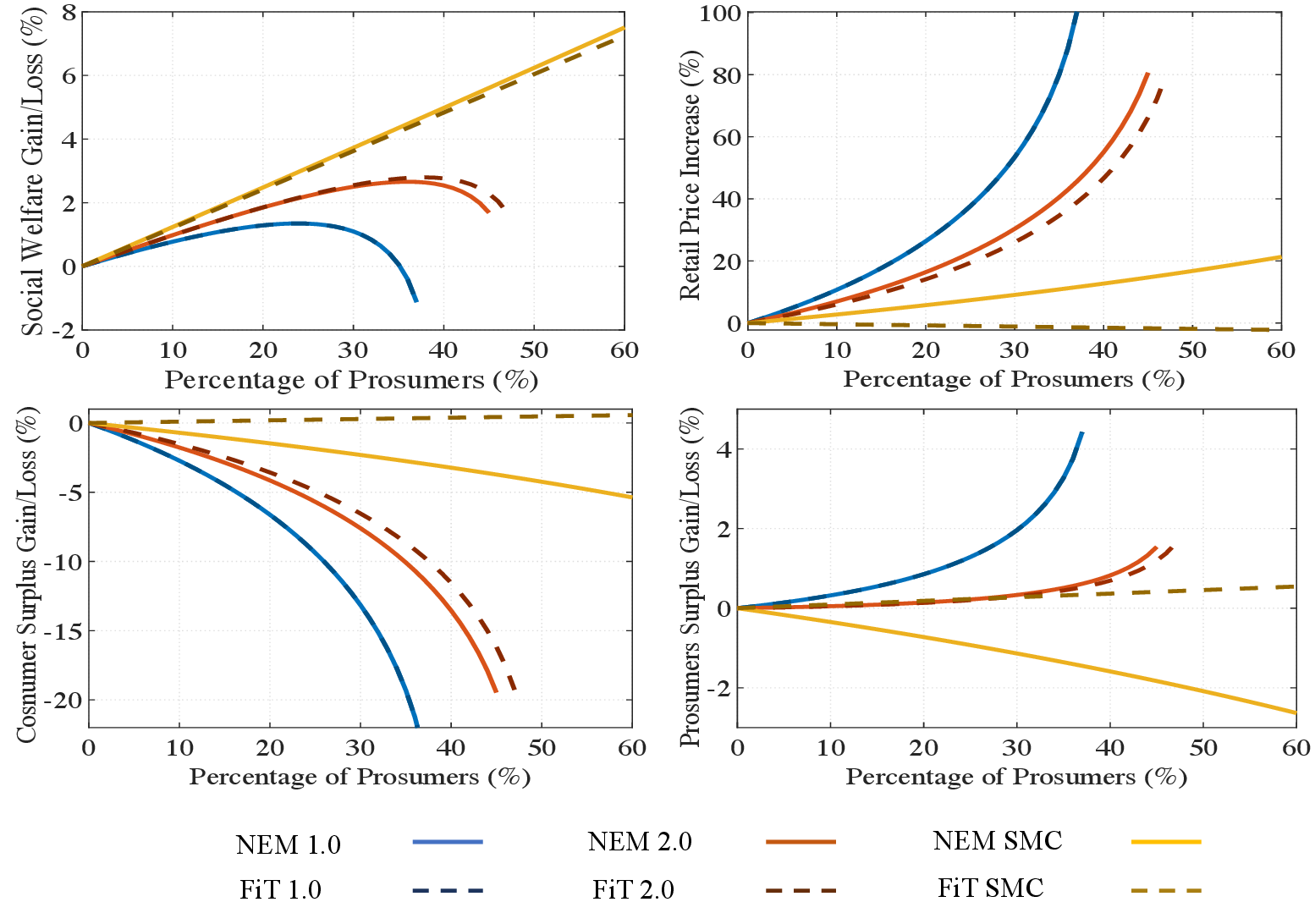}
    \caption{NEM X and FiT X prosumers. Clockwise from top-left: social welfare, retail price, prosumer surplus, and consumer surplus percentage changes.}
    \label{fig:NEMFiTNum}
\end{figure}

\subsubsection{Social welfare}\label{sec:socialwelfareNum}
Consider first the welfare effects of NEM/FiT 1.0-2.0 policies.  The top-left panel of Fig.~\ref{fig:NEMFiTNum} shows that the percentage change of the short-run social welfare (over the cases with 0\% adoption rate) versus the prosumer population were all increasing when the prosumer population was low and decreasing when the prosumer population was high. The somewhat surprising decline of social welfare at high prosumer population was due to the breakeven requirement of the retail tariff as discussed in Sec.~\ref{sec:adoptionModelDynamics}, and verified by the rest of Fig.~\ref{fig:NEMFiTNum}.

In particular, the top-right panel shows that NEM/FiT 1.0--2.0 policies have accelerated growing retail prices ($\pi^+$), resulting in precipitous declining consumer surpluses shown in the bottom left panel.  Although the prosumer surplus (bottom-right panel) and the environmental benefits grew, the combined effect was that the weight of increasing retail price eventually dragged down the overall social welfare. The retail price increase under NEM 2.0 and NEM SMC is faster than it is under FiT 2.0 and FiT SMC since export rate reductions affect the bill savings of FiT X prosumers more profoundly compared to NEM X prosumers, which enables the utility to breakeven at a lower $\pi^+$.

In contrast, NEM/FiT SMC policies exhibited different characteristics with growing social welfare with increasing prosumer populations.  The SMC pricing of generation eliminates the disadvantage of procuring renewables from DER generation in the distribution system, and the growth of environmental benefits lifted the overall social welfare.

\subsubsection{Market potential}
The market potential was calculated using (\ref{eq:MktPoten}), which depends on the expected payback time solution in (\ref{eq:TNPpaybackTime}). Table \ref{tab:NEMFiTNumMktPoten} shows the market potential under the studied NEM X and FiT X policies. Similar to Sec.~\ref{sec:socialwelfareNum}, when $\pi^-=\pi^+$, NEM 1.0 and FiT 1.0 policies were equivalent. The mild export rate reduction affected FiT 2.0 prosumers more than NEM 2.0 prosumers because the whole generation under FiT 2.0 was compensated at $\pi^-$, whereas only the net-exported energy faced $\pi^-$ in NEM 2.0. Therefore, the market potential under NEM 2.0 was higher. The same reasoning applies to the NEM SMC and FiT SMC curves, although both policies significantly increased the DER payback time, which stalled the market adoption potential. Lastly, from Table~\ref{tab:NEMFiTNumMktPoten}, lowering $\pi^-$, delays the occurrence of utility death spirals, under which a $\pi^+$ achieving the revenue adequacy constraint in (\ref{eq:Ramseymaximization}) is infeasible.

\begin{table}[htbp]
\caption{Market potential of NEM X and FiT prosumers.}
\label{tab:NEMFiTNumMktPoten}
\centering
\resizebox{0.47\textwidth}{!}{%
\begin{tabular}{@{}cccccccccc@{}}
\toprule \toprule
Policy  & \multicolumn{9}{c}{Percentage of Prosumers (\%)}                           \\ \midrule
--      & 0     & 10     & 20     & 30     & 35     & 40    & 45     & 50    & 60    \\ \midrule
NEM 1.0 & 14\%  & 17.1\% & 21.9\% & 29.1\% & 35.1\% & --    & --     & --    & --    \\
FiT 1.0 & 14\%  & 17.1\% & 21.9\% & 29.1\% & 35.1\% & --    & --     & --    & --    \\
NEM 2.0 & 6.4\% & 8.3\%  & 11.1\% & 15.1\% & 18\%   & 22\%  & 28.1\% & --    & --    \\
FiT 2.0 & 4.6\% & 6.2\%  & 8.5\%  & 12\%   & 14.6\% & 18\%  & 23.3\% & --    & --    \\
NEM SMC & 0.6\% & 0.7\%  & 0.8\%  & 0.9\%  & 1\%    & 1.1\% & 1.2\%  & 1.3\% & 1.5\% \\
FiT SMC & 0\%   & 0\%    & 0\%    & 0\%    & 0\%    & 0\%   & 0\%    & 0\%   & 0\%   \\ \bottomrule \bottomrule
\end{tabular}%
}
\end{table}

\subsection{Long-run performance of NEM X policies}
We explored the long-run effect of the NEM X policies in Table \ref{tab:PlotCases} in addition to NEM 1.0 and NEM 2.0. The policies NEM D\# are dynamic, with changing tariff parameters every year. NEM D1, NEM D2, and NEM D3 annually vary $\pi^-$, whereas NEM D4 annually varies $\pi^0$ uniformly on all customers. NEM 1.0, NEM 2.0, and NEM 3.0 policies are all static, meaning that their tariff parameters do not change over time.
The exogenous parameters $\xi$ and $\theta$ are annually varied as follows: a) the PV installation cost $\xi$ decreases by 3.5\% each year starting from $\xi_0$, and b) the expected utility fixed cost $\theta$ increases by 2.6\% each year starting from $\theta_{\mbox{\tiny PGE}}$, which are given in Appendix~\ref{sec:appendixNumData}.

\begin{table}[htbp]
\caption{Description of long-run case studies.}
\label{tab:PlotCases}
\centering
\resizebox{0.45\textwidth}{!}{%
\begin{tabular}{@{}cccccc@{}}
\toprule \toprule
Policy     & Tariff   & Initial $\pi^-$          & Final $\pi^-$            & \begin{tabular}[c]{@{}c@{}}Annual\\ decrement\end{tabular} & \begin{tabular}[c]{@{}c@{}}Fixed\\ charges\end{tabular} \\ \midrule
NEM D1 & one-part & $\pi^+$                  & $0.4\pi^+$               & 2.4\%                                                      & --                                                      \\
NEM D2 & one-part & $\pi^+$                  & $0.3\pi^+$               & 2.8\%                                                      & --                                                      \\
NEM D3 & one-part & $\pi^+$                  & $0.25\pi^+$              & 3\%                                                        & --                                                      \\
NEM D4 & two-part & $\pi^+-0.035$            & $\pi^+-0.035$            & 0\%                                                        & 0-40$^\dagger$                                                  \\
NEM 3.0    & two-part & $\pi^{\mbox{\tiny AVR}^\ast}$ & $\pi^{\mbox{\tiny AVR}}$ & 0\%                                                        & 8$^\ddagger$                                                   \\ \bottomrule \bottomrule
\end{tabular}%
}
\footnotesize{ \begin{flushleft} \vspace{-0.05cm}
\hspace{0.25cm}$^\ast$ The Californian avoided cost rate $\pi^{\mbox{\tiny AVR}}$ is adopted from \cite{E3_AvoidedCost}.\\
\hspace{0.25cm}$^\dagger$ Uniform connection charges (\$/month). The increment is \$2/year.\\
\hspace{0.25cm}$^\ddagger$ CBC charges \$/kWDC/month. The value is taken from \cite{CPUC_NEM3}.
 \end{flushleft}}
\end{table}

\subsubsection{Market adoption}
Figure \ref{fig:adoption_policies} shows that NEM~1.0 and 2.0 fostered an accelerated growth of the percentage of prosumers, which created upward price pressure for the utility to cover the loss of revenue. The two policies could not cover their costs because of the death spiral phenomenon. With the significantly reduced compensation rate and a fixed CBC charge of \$40/month on a 5kW system,  the NEM 3.0 stalled the potential of the DER adoption. The long-run adoption of dynamic NEM D\# policies exhibited quite different characteristics. NEM D4, which dynamically increases the non-volumetric connection charges, yielded a lower adoption in the early stages compared to NEM D1-D3. The reason was that the connection charges are taken from all customers (consumers and prosumers), whereas the effect of export rate reduction in NEM D1-D3 comes only from the prosumers $\gamma$. The \$2/year increase in connection charges under NEM D4 was not high enough to recover the utility's lost energy sales due to increased adoption, which led to successive retail price increases, eventually leading to a death spiral. The case was different under NEM D1-D3. Decreasing the ratio $\pi^-$ to $\pi^+$ at a rate of 2.8\% and 3\% stabilized the long-run adoption, but they resulted in a lower percentage of prosumers, due to prolonged DER investment payback times. The reduction of $\pi^-$ in NEM D1 prevented the occurrence of a death spiral and ushered the adoption, but at the cost of higher cost-shifts, as discussed in the next section. The dynamic policies show that, from a DER adoption perspective, reducing the export rate is more effective and adoption-sustaining than increasing the uniform fixed charges.

\subsubsection{Cost shifts}
\par
\begin{figure}[htbp]
    \centering
    \includegraphics[scale=0.35]{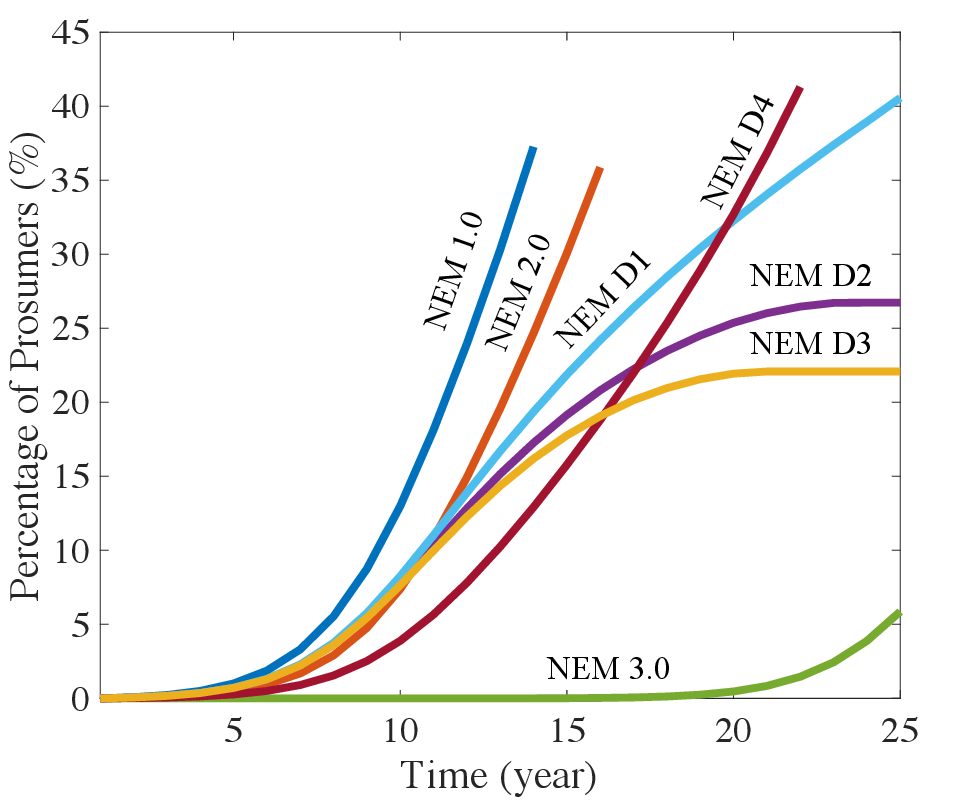}
    \caption{Long-run adoption of NEM X policies.}
    \label{fig:adoption_policies}
\end{figure}

 The cost-shifts of the tested NEM X policies are presented in Table \ref{tab:costshift_policies}. Although NEM 1.0 and NEM 2.0 accelerated DER adoptions, as shown in Fig.\ref{fig:adoption_policies}, the resulting long-run cost-shifts under them were quite high compared to other policies, which is a result of the high price markup between the retail and the utility avoided cost rates. Note that NEM D2-D3 policies were more effective in suppressing cost-shifts than NEM D1 and NEM D4. In fact, NEM D4 was effective in the short-run cost-shifts reduction but failed to reduce cost-shifts in the long-run.
The effect of $\pi^-$ reductions in NEM D2-D3 mounts as the adoption rate grows. 
With decreasing $\pi^-$ under NEM D2-D3, the growth rate of cost-shifts slowed with the increasing prosumer fraction and eventually declined. This was because the effect of $\pi^-$ reductions on prosumers' bill savings dominated their benefit of $\pi^+$ increase. Lastly, NEM 3.0 significantly reduced and almost eliminated cost-shifts, but at the cost of stalled rooftop PV adoption, as shown in Fig.\ref{fig:adoption_policies}.

\begin{table}[htbp]
\caption{Long-run cost-shifts (\$/customer/month) of NEM X policies.}
\label{tab:costshift_policies}
\centering
\resizebox{0.37\textwidth}{!}{%
\begin{tabular}{@{}cccccccccc@{}}
\toprule \toprule
Policy  & \multicolumn{8}{c}{Time (year)}                           \\ \midrule
--      & 3 &    6 &  9  & 12  & 15  & 18 & 21 & 24    \\\midrule
NEM 1.0 &      0.6 & 4.5 & 20.3 & 63.4 & -- & -- & -- & --     \\
NEM 2.0 &      0.1 & 1 & 5.3 & 18.8 & 50.4 & -- & -- & --    \\
NEM D1 & 0.4 & 2.4 & 8.8 & 18.7 & 28.9 & 38.3 & 46.7 & 55.1  \\
NEM D2 & 0.4 & 2.3 & 8.0 & 15.6 & 21.6 & 24.6 & 24.5 & 21.3  \\
NEM D3 & 0.4 & 2.2 & 7.5 & 14.1 & 18.4 & 19.3 & 17.2 & 13.7   \\
NEM D4 & 0.1 & 0.7 & 3.3 & 9.5 & 20.3 & 38.3 & 75.8 & --   \\
NEM 3.0    & 0 & 0 & 0 & 0 & 0 & 0.1 & 0.6 & 2.8  \\
\bottomrule \bottomrule
\end{tabular}%
}
\end{table} 

\section{Literature on NEM and FiT policy analysis}
\label{sec:appendixLR}
This section presents a non-exhaustive review of related work on the analysis of NEM and FiT policies with respect to 1) social welfare, 2) customer equity, and 3) BTM DER adoption.

\subsection{Social welfare}
The classical theory about the social welfare of public utility pricing focuses on pricing efficiency defined by the constrained social-welfare maximization of Boiteux-Ramsey \cite{Brown&Sibley:1986book}. Yamamoto gave perhaps the earliest social welfare analysis of NEM and made comparisons among NEM, FiT, and net purchase and sale (a special case of NEM as defined in this paper) \cite{YamamotoPricingEF:12SE},  where explicit characterizations of consumer/prosumer surpluses were provided.  Yamamoto's model takes into account the differentiated retail and export rates as in NEM 2.0, although the two rates are linked through the cost of DER installation, rather than determined simultaneously by the regulator via the Boiteux-Ramsey pricing in this work.  A key difference between Yamamoto's approach and that presented in \cite{Alahmed_Tong:22IEEETSG} is that the latter characterizes the consumption as an implicit function of the available level of DER, whereas Yamamoto's approach used the correlation coefficients between the export rate and the level of DER, which is quite difficult to estimate in practice. Because of this difference, the household consumption in \cite{YamamotoPricingEF:12SE} decreases when the consumer becomes a prosumer, opposite to that shown in \cite{Alahmed_Tong:22IEEETSG}. A broader definition of social welfare as the sum of customers’ surplus, utility profit, and environmental benefits positive externality is used in \cite{Alahmed_Tong:22IEEETSG, YamamotoPricingEF:12SE} and this work. They all show 1) a decreasing consumer surplus after introducing the FiT and NEM programs, 2) an equivalent NEM X and FiT X social welfare when $\pi^- = \pi^+$, 3) a higher retail price under NEM X compared to FiT X, and 4) and an increasing prosumer surplus at the cost of consumers' surplus loss, when the export rate is close to the retail price. The work in \cite{Alahmed_Tong:22IEEETSG} shows that bringing the export rate of NEM X closer to the avoided cost rate gives the highest welfare, as the retail rate becomes relatively closer to the SMC rate.

\par The impact of designing the NEM export rate is analyzed in \cite{Brown&Sappington:17EJ,YAMAMOTO2017312,Boero_DERmicroeconomics:16SE}. The authors in \cite{Brown&Sappington:17EJ} find that, in addition to negative distributional effects, compensating excess BTM generation at the retail rate creates considerable welfare losses. Additionally, they concluded that, when the fixed costs of managing the transmission and distribution systems are high, it is optimal to set the retail rate higher than the export rate ($\pi^+\geq \pi^-$). The opposite is true ($\pi^- \geq \pi^+$) when the fixed costs are negligible and the marginal cost of centralized production is high.
The welfare implications of two-part tariff\footnote{Two-part tariffs consist of a volumetric charge (the price of a kWh of energy consumed) and a lump sum fee (e.g. consumption-independent connection charges) \cite{WalterOi_Disneyland:71QJE}.} NEM and FiT models are investigated in \cite{YAMAMOTO2017312}, where the welfare under FiT is shown to be higher than under NEM if the export rate is equivalent to the avoided cost rate. More interestingly, the author showed that the export rate that maximizes social welfare is very close to the retail rate before applying the NEM or FiT programs (i.e. when there is no market adoption). This is consistent with the maximum social welfare achieved under NEM SMC and FiT SMC, which both maintained an export rate that is close to the retail rate when there is no adoption. 

\par The consumer and prosumer surplus tradeoffs are analyzed under both NEM X and FiT X in \cite{Boero_DERmicroeconomics:16SE} and under NEM X in \cite{Gautier_NEMXDeplymentandEquity:18JRE}. Through a theoretical model that maximizes the social welfare subject to a utility revenue adequacy constraint under inelastic customers, the authors in \cite{Boero_DERmicroeconomics:16SE} show that, unless exogenous market shocks such as lower PV installation cost or higher environmental benefits are present, the surplus transfer from consumers to prosumers is the only mechanism to encourage and sustain PV adoption. Moreover, the authors conclude that, compared to NEM, higher PV adoption under FiT underlies a more negative impact on consumer surplus because the retail rate needs to be higher. The surplus transfer from consumers to prosumers is also discussed in \cite{Gautier_NEMXDeplymentandEquity:18JRE} and proven to be more severe under NEM 1.0 compared to NEM X policies with $\pi^-<\pi^+$ and shorter net-billing periods. This is partially due to the reduced amount of self-consumed DER under NEM 1.0, which places more burden on the utility cost recovery \cite{Borenstein&Bushnell_DoTwoElectricityPrices:18NBER} in addition to introducing reliability issues \cite{RATNAM:15AE}. The conclusion of \cite{Boero_DERmicroeconomics:16SE,Gautier_NEMXDeplymentandEquity:18JRE} corroborates the results of the consumer and prosumer surplus transfer in this work.

\par A narrower definition of social welfare is considered in \cite{Daniel_PartI:18TPS, AlvarezConnectionChargesPartII:18TPS,Tong_DERdynamics:20TAC} as the sum of utility and customer population surpluses. The work in \cite{Daniel_PartI:18TPS, AlvarezConnectionChargesPartII:18TPS} establishes a stochastic framework to study the impact of applying NEM 1.0 with a two-part tariff under flat and dynamic rates on social welfare. The authors show that, with NEM 1.0, the customer surplus decreases with the level of DER integration due to the increasing retail rate, as the utility struggles to recover the operating costs through energy sales. However, a breakdown of the welfare to prosumers and consumers surpluses is not presented, and the environmental impact of solar adoption is not incorporated. The work in \cite{Tong_DERdynamics:20TAC} shows that higher connection charges yield higher long-run welfare, as connection charges reduce the utility burden of fixed cost recovery. The environmental benefits of solar are also ignored in \cite{Singh&SchellerWolf:22Informs}, under which the authors examine the effect of tariff structures on social welfare, defined as the sum of surpluses of the utility, the solar companies, and the price-inelastic customers. They find that socially optimal NEM 1.0 tariff models must include both volumetric discrimination, and customer technology discrimination, where prosumers and consumers face different tariff structures.
\par The work in \cite{ANSARIN2020115317} studies the social welfare of three volumetric rate variants and a tariff with demand charges under NEM X. Characterized by the deadweight loss in customer surplus, it is shown that the social welfare under IBR and flat rates are worse than under TOU and demand charges-based rates, especially as the adoption rate increases.

\par The approach presented here follows that developed in \cite{Daniel_PartI:18TPS,AlvarezConnectionChargesPartII:18TPS,Tong_DERdynamics:20TAC} in analyzing NEM 1.0 and \cite{Alahmed_Tong:22IEEETSG} for NEM X. In particular, this work generalizes the NEM 1.0 long-run dynamics of rate-setting process model in \cite{Tong_DERdynamics:20TAC} to NEM X, while using a broader definition of social welfare compared to \cite{Daniel_PartI:18TPS,AlvarezConnectionChargesPartII:18TPS,Tong_DERdynamics:20TAC}, by considering the environmental benefit externality and the distinction of prosumers and consumers surpluses.

\subsection{Equity considerations} 

In addition to the economic efficiency principle in retail rate design, equity consideration plays an important role in the rate-making process \cite{bonbrightprinciples:1961}.
Several equity concerns have been raised recently. Researchers argued that, under NEM X, the relatively high installation cost of BTM DER and lack of incentives for low-income customers to adopt DERs create bill gaps between prosumers and consumers, consumers, which include disproportionately larger less affluent consumers \cite{Next10Report,Nelson_IncomeGap:11EAP}. The authors in \cite{BorensteinPrivateNetBenefits:15NBER,DARGHOUTH20115243}, for example, find that the application of IBR under NEM 1.0 in California contributes to the surge in the DG adoption levels by customers within higher income brackets, and enables them to achieve significant bill savings by avoiding the high-priced blocks. This has pushed some PUCs to consider policies that mandate TOU rates \cite{cpuc_NEM2} or even levy discriminatory charges on adopters \cite{CPUC_NEM3}. Moreover, although a few papers claim that treating DER adopters as a separate customer class (retail rate differentiation) is needed to attain societal and rate efficiency objectives \cite{Singh&SchellerWolf:22Informs}, many state commissions have rejected proposals differentiating adopters and non-adopters \cite{StantonNEMreview:19NARUC}. 

\par In addition to inequities resulting from the income gap between DER adopters and non-adopters, the most well-known equity issue under NEM X is the cross-subsidies resulting from prosumers shifting cost recovery obligations to consumers. As analyzed in the seminal work of Faulhaber in \cite{Faulhaber:75}, such cross-subsidies are the result of part of customers (prosumer in NEM case) not paying their "fair share" of common costs.

\par The gap between the wholesale price of electricity and the compensation price $\pi^-$ for excess DER generation creates a revenue loss, which is socialized through retail rate increase \cite{Borenstein_CostShifts:21HaaSBlog}. Effectively, DER adopters shift some of the utility fixed and capacity costs to non-adopters. The authors in \cite{Next10Report} attribute this cross-subsidy to the regressiveness of volumetric charge recovery\footnote{volumetric charge recovery is the reliance on a tariff that charges customers solely on the amount they consume to recover utility's costs.}, which inflated the retail rate in California for example to 2-3 times the SMC rate of electricity. Pure volumetric tariffs are also identified in \cite{PicciarielloSubsidyQuantify:15UP} as an inadequate tariff design that does not reflect the household-driven costs. The shifted costs proliferate under an IBR-based volumetric cost recovery \cite{UtilityRegulatoryBusinessModel:16NREL}, which is proven to achieve unduly bill savings for prosumers \cite{BorensteinPrivateNetBenefits:15NBER}.

\par Furthermore, as investigated in \cite{PicciarielloSubsidyQuantify:15UP,SergiciSubsidiesQuantify:19EJ,EID_CostRecoveryShifts:14EP}, such cross-subsidies intensify as the adoption rates grows under NEM 1.0. Although NEM 1.0 is more lucrative to prosumers compared to NEM 2.0 and beyond, the benefits to prosumers in NEM 1.0 mostly comes at the price of creating more cost
shifts to consumers \cite{FutureofSolar_MIT:15MITreport}. On the other side, DER proponents argue that the price differential between the retail and export rates should not be large when considering DER brings added benefits to the utility such as avoiding distribution system upgrades, losses, and environmental taxes \cite{LucasDavis_CostShifts:18Haas}. Some researchers argue, however, that the avoided utility cost due to BTM DER is minimal \cite{COHEN2016139}.
\par To partially address cost shifts, some PUCs revised the export rate to accurately reflect the value DER adds to the grid. The effects of export rate reduction on reducing cost-shifts are investigated in \cite{Alahmed_Tong:22IEEETSG}.
Furthermore, to suppress cross-subsidies, some researchers propose income-based fixed charges, which, specifically, target wealthier customers who are more inclined to adopt BTM DER and with relatively larger capacities \cite{Next10Report}. The role of fixed charges under NEM in preventing death spirals and reducing subsidies is also empirically shown and emphasized in \cite{Clastres_CrossSubsidies:19UP,Alahmed_Tong:22IEEETSG,EdisonReport_Borlick_Wood}.
\par With feed-in metering that measures separately gross consumption and DER, FiT can be structured to eliminate cost shifts by accurately accounting for the actual power consumption and power generation \cite{FeedInCostShifts:16Brookings}. Thus, a careful design of the export rates is required.

\subsection{DER adoption and long-run performance} The revenue metering arrangement and the corresponding retail tariff design have the potential of ushering or stalling BTM DER adoption. The retail policy effect on rooftop solar adoption was analyzed in \cite{Alahmed_Tong:22IEEETSG}, which leverages a characterized DER-elastic consumer decision model on the regulator decision problem, and shows the efficiency of some adoption controlling tools such as export rate reduction, dynamic pricing, and fixed charge increases. The authors show that, under NEM 1.0 and NEM 2.0, the DER payback time is reasonably short compared to an avoided-cost compensating NEM X, resulting in a rapid DER installation growth. Similarly, it is concluded in \cite{COMELLO201746} that the payback time is substantially prolonged if the export rate is reduced. The authors argue that, as long as the export rate is above the levelized cost of electricity, the adoption of rooftop solar would continue. This conclusion, however, is based on a demand that is inelastic to the export rate reductions.

\par Furthermore, in \cite{DARGHOUTH20115243}, the effect of metering, and rate design on prosumers' bill savings, which is a crucial factor in adoption patterns, is evaluated. The paper shows that, under IBR, and as the PV generation increases, the incremental value of bill savings decreases, as the net consumption faces a declining marginal price. This shows that high-usage customers disproportionately benefit from IBR. The net-billing period effect is also studied in \cite{DARGHOUTH20115243}, where it is shown that shorter net-billing periods increase the amount of exported generation, which negatively influences bill savings if the export rate is below the retail rate.
\par Moreover, a well-structured analytical framework to study the PV adoption process under NEM 1.0 as a nonlinear dynamical system was derived in \cite{Tong_DERdynamics:20TAC}. The paper shows the effectiveness of uniform fixed charges in always ensuring a stable rooftop solar adoption, under any given utility fixed cost. Uniform fixed charges, however, have been criticized for creating inequities between low consuming customers, who are usually low-income, and high-income customers \cite{Borenstein_FixedCost:16EJ}. The potential of death-spiral associated with the uncontrolled adoption rates under NEM 1.0 has been investigated in \cite{CaiRatesImpact:13EP}. The authors show that, due to bill reductions, the increasing retail rate shortens the time of the installed PV capacity to reach 15\% of the peak demand by almost 4 months. Also on NEM 1.0, the papers in \cite{denholm_drury_margolis_2009,Darghouth_NEMFeedbackloops:16AE} study the scenarios of higher DER penetration under NEM 1.0 and find that the struggle of utilities to recover their costs induces potential of death spirals. 

\par In addition to solar-only adoption under NEM X, the economic feasibility of solar+storage packages under different rate designs including NEM 2.0 is empirically studied in \cite{Darghouth_PVSolar:19LBNL}. The authors find that the price differential $\pi^+-\pi^-$ and the TOU rate parameters (e.g. peak period and peak ratio) heavily influence the bill savings achieved by solar+storage, and therefore their adoption. The literature on the adoption of solar and storage packages, however, is rather scarce.

\par The DER adoption under FiT has also been studied. An analysis of the effect of differentiating residential customers FiT's export rate in Germany based on their installed capacity on the adoption of solar is presented in \cite{GermeshausenFiT:19ZEW}. The author finds that reducing the export rate by 5\% for larger-scale residential PV leads to a 29\% reduction of newly installed PV capacities. Also, in \cite{Singh&SchellerWolf:22Informs}, a FiT design that results in a stable, socially optimal adoption, and with no cross-subsidies, but at the cost of discriminating consumers and prosumers is proposed. However, using a non-discriminatory tariff, the work in \cite{YAMAMOTO2017312} finds that a FiT policy with an export rate equivalent to the avoided cost rate can maximize the social welfare and promote the DER adoption, provided that a uniform two-part tariff is implemented as an adoption pace-controlling tool.  
The authors in \cite{Varaiya_NEMA:19TSG} analyze the economic feasibility of community solar under different revenue metering structures. They show that, unlike FiT, NEM X policies make community solar favorable for adopters. Lastly, in \cite{BabichLobelYucel:20MSOM}, the government option of considering FiT or NEM 1.0 with tax rebates on rooftop solar adoption is compared. The government's goal is to promote solar while maximizing the expected difference between the societal benefits of installing solar and the cost of the subsidies provided over time. The researchers found that the government favors FiT if the retail prices are highly volatile and the investment cost is stable.

\section{Conclusion}
\label{sec:conclusion}
The retail electricity market is at an inflection point, facing significant changes in the existing NEM tariff policies in addressing concerns of rising energy costs, equity, and long-term sustainable growth of BTM solar. To this end, a characterization of prosumer and consumer behaviors in response to various NEM policy designs is a first step to gaining insights into short and long-run impacts of potential NEM tariff evolution. This paper contributes to this objective through an analytical approach with an inclusive NEM model that captures different generations of NEM policies. The resulting structure of the optimal consumption policy for prosumers and consumers makes it possible to evaluate the impacts of NEM tariff designs.

Although the modeling, optimization, and microeconomic analysis are couched in a setting for residential customers, the insights gained from this work broadly apply to energy management problems for commercial and industrial customers.  We summarized these insights from two perspectives.  

The engineering perspective is that the evolving NEM X policies bring significant engineering challenges and opportunities in demand-side energy management in a retail market with substantial distributed energy resources.  The differentiated import-export pricing in NEM X heightens the need for active energy management and adaptive control strategies that are elastic to possibly dynamic NEM pricing policies and stochastic BTM generations.  Such active energy management strategies result in economic benefits to prosumers and reliability benefits to system operations by reducing reverse and peak power flow. Quantifying the latter requires additional research that considers jointly the control and optimization challenges of the system operator and that of the prosumers studied in this work.

The economic perspective is that the parametric model of NEM/FiT X provides an analytical framework to characterize the short-run social welfare distribution (among consumers and prosumers), environmental externalities, and market potential of DER adoption.  Such characterizations reveal ways of setting NEM parameters to increase short-run social welfare, reduce cross-subsidies for prosumers by consumers, and mitigate the negative impacts of significantly reduced export rate ($\pi^-$) in NEM X parameters on the growth of BTM DER.

\section*{Acknowledgments}
The authors would like to thank helpful comments from Prof. Steven
Low. This work was supported in part by National Science Foundation under Awards No.:~1809830 and No.:~1932501.

{
\bibliographystyle{IEEEtran}
\bibliography{AlahmedTong_EIR}
}

\section{Appendices}

\subsection{Notations and nomenclature}
\label{sec:appendixNotations}
We use boldface for vectors and matrices.
Key designations of symbols are given in Table \ref{tab:symbols}.

 {\small
\begin{table}[htbp]
\caption{\small Major designated symbols (alphabetically ordered).}\label{tab:symbols}
\begin{center}
\vspace{-1em}
\begin{tabular}{|ll|}
\hline\hline
${\bf 0}$: & vector of all zeros.\\
$C$: & utility cost function.\\
$\chi(\cdot)$: & indicator function.\\
$\bm{d}$: & consumption bundle in a net-billing period.  \\
$\bar{\bm{d}}$: & consumption bundle upper limit.  \\
$\bm{d}^\ast$: & the optimal consumption bundle.\\
$\tilde{d}_k$: & exercised consumption decisions sum up to $k-1$.\\
$d^\pi_{t,i}$: & device $i$ MPC-based optimal consumption.\\
$d^+_i, d^-_i, d^o_i $: & the three zones optimal consumptions of device $i$.\\
$d^+, d^-$: & thresholds of the optimal prosumer policy.\\
$\Delta P_\pi$: & bill savings due to BTM DER under tariff $\pi$.\\
$\mathcal{E}$: &  environmental and health benefits of BTM DER.\\
$\eta, \eta_\infty$: & sigmoid function and market potential.\\
$\gamma$: & the fraction of prosumers in the population.\\
$h$: & adoption model.\\ 
$J$: & num. of net-billing periods in a rate-setting period.\\
$M$: & number of devices.\\
$\nu$: & BTM DER system degradation factor.\\
$P_{\pi}^{\mbox{\tiny NEM}},P_{\pi}^{\mbox{\tiny FiT}}$: &  NEM and FiT payments under tariff $\pi$.\\
$\pi=(\pi^+,\pi^-\pi^0)$: & tariff (buy rate, export rate, fixed charge). \\
$\pi^{\mbox{\tiny AVR}}, \pi^{\mbox{\tiny SMC}}, \pi^e$: & AVR, SMC, and environmental prices.\\
$\psi^\pi_\gamma$:& cost-shifts under tariff $\pi$ and adoption level $\gamma$.\\
$r$: & BTM DER in a net-billing period.  \\
$\hat{r}_k$: & BTM DER forecast at decision period $k$.\\
$\tilde{r}_k$:& realized BTM DER up to $k-1$.\\
$\mathbb{R}^M,\mathbb{R}_+^M$ & sets of $M$ dim. real and positive real vectors.\\
$s$: & DER adoption curve.\\
$S^\pi$: & customer surplus under $\pi$.\\
$S^\pi_c, S^\pi_u$: & customer and utility surpluses under $\pi$.\\
$T$: & number of decision periods in a net-billing period. \\
$T_{pb}^\pi$: & DER payback time under tariff $\pi$.\\
$\theta_n$: & utility's expected fixed cost in period $n+1$.\\
$U(\cdot), U_i (\cdot)$: & utility functions\\
$\tilde{U}_k$: & aggregate utility of consumption up to $k-1$.\\
$V(\cdot), V_i (\cdot)$: & marginal utility functions\\
$W^\pi_\gamma$:& social welfare under tariff $\pi$ and adoption level $\gamma$.\\
$\xi$: & DER installation cost.\\
$y$: & net energy consumption of both customer classes.\\
$z$: & net energy consumption in a net-billing period.\\
$\zeta$: & interest rate.\\
\hline\hline
\end{tabular}
\end{center}
\end{table}
} 

\subsection{Theorems and proofs}
\label{sec:appendixProofs}
\begin{theorem}[Prosumer decision under NEM X \cite{Alahmed_Tong:22IEEETSG}]  \label{thm:structure}
Given the NEM X parameter $\pi=(\pi^+,\pi^-,\pi^0)$ and the marginal utilities vector $(V_1,\cdots, V_M):=\frac{\partial U(\bm{d})}{\partial \bm{d}}$ of consumption devices, under the assumptions of strictly concave utility, $\pi^+>\pi^-$, non-binding budget constraint and non-degeneracy condition of (\ref{eq:prosumerOPT_NEM}), the optimal prosumer consumption policy is given by two thresholds
\begin{equation}
\begin{array}{l}
d^+:=\sum_{i=1}^M  \max \{0,\min\{V_i^{-1}(\pi^+),\bar{d}_i\}\},\\
d^-:=\sum_{i=1}^M \max  \{0,\min\{V_i^{-1}(\pi^-),\bar{d}_i\}\} \ge d^+\\
\end{array}
\end{equation}
that partition the range of DER production into three zones:
\begin{enumerate}
\item \underline{Net consumption zone:  $r<d^+$}. The prosumer is a net-consumer with consumption
\begin{equation} \label{eq:d_i^+}
d^+_i =\max \{0,\min\{V_i^{-1}(\pi^+),\bar{d}_i\}\} \ge 0,~~\forall i.
\end{equation}
\item  \underline{Net production zone:  $r>d^-$}. The prosumer is a net-producer with consumption
\begin{equation} \label{eq:d_i^-}
d^-_i=\max  \{0,\min\{V_i^{-1}(\pi^-),\bar{d}_i\}\} \ge d_i^+,~~\forall i.
\end{equation}
\item  \underline{Net-zero energy zone:  $d^+ \le r \le d^-$}. The prosumer is a net-zero consumer with consumption:
\begin{equation} \label{eq:d_i^o}
 d^o_i(r)  =  \max\{0, \min\{V_i^{-1}(\mu^*(r)),\bar{d}_i\}\} \in [d^+_i, d^-_i], \forall i
\end{equation}
where $\mu^*(r) \in [\pi^-, \pi^+]$ is a solution of
\begin{equation}
\sum_{i=1}^M \max\{0, \min\{V_i^{-1}(\mu),\bar{d}_i\}\} = r.
\label{eq:r}
\end{equation}
Furthermore, $d_i^o(\cdot)$ is continuous and monotonically increasing in $[d^+_i,d^-_i]$.
\end{enumerate}
\end{theorem}
{\em Proof:} See \cite{Alahmed_Tong:22IEEETSG}. \hfill $\blacksquare$

\begin{theorem}[Adaptive scheduling of consumptions]
\label{thm:MPCversionofOCT}
\par Given the NEM X parameter $\pi$ and marginal utilities $V_{t,i} \in \mathbb{R}^T$, under the concavity and temporal additivity of $U(\cdot)$, in addition to having $\pi^+\geq \pi^-$, the adaptive consumption decision at each $k$ measurement period is given by two thresholds:
\begin{equation}
\label{eq:thresholdsMPC}
\begin{array}{l}
d^+_k:= \tilde{d}_k+\sum_{i=1}^M \sum_{t=k}^T d_{t,i}^+,\\
d^-_k:=  \tilde{d}_k+\sum_{i=1}^M \sum_{t=k}^T d_{t,i}^- \ge d^{+}_k,\\
\end{array}
\end{equation}
where $\tilde{d}_{k}:= \sum_i^M \sum_{t=1}^{k-1} d_{t,i}^\pi$, is the sum of exercised consumption at time $t=k$. The $k$th consumption decision is:
 \begin{equation}
     d^\pi_{t,i}=\begin{cases}
 d^+_{t,i}& \text{ if } \tilde{r}_{k}+\sum_{t=k}^T \hat{r}_t \leq  d^+_k\\ 
 d^-_{t,i}& \text{ if } \tilde{r}_{k}+\sum_{t=k}^T \hat{r}_t \leq  d^-_k\\
 d^o_{t,i} & \tilde{r}_{k}+\sum_{t=k}^T \hat{r}_t \in \left[d^+_k , d^-_k\right],
\end{cases}
 \end{equation}
 where $\tilde{r}_{k} := \sum_{t=1}^{k-1} r_t$, and for every $i$, $d^o_{t,i}$ is the $k$th element of $\bm{d}_i^o(\bm{r})  =  \max\{\bm{0}, \min \{V_i^{-1}(\bm{1}\mu^*(\bm{r})),\bar{\bm{d}}_i\}\}$, where $\mu^\ast(\cdot)\in [\pi^-,\pi^+]$ is a solution of:
 \begin{equation}
 \label{eq:equalityAdaptive}
\tilde{d}_{k}+\sum_{i=1}^M \sum_{t=k}^T \max{(0,\min{(V^{-1}_{t,i}(\mu),\bar{d}_{t,i})})} = \tilde{r}_{k}+\sum_{t=k}^T  \hat{r}_t.
\end{equation}
 \end{theorem}

{\em Proof of Theorem \ref{thm:MPCversionofOCT}:}
Recall the $T$ sensing and control intervals. Assume that the consumer already exercised $k-1$ consumption decisions out of the $T$ available ones, with:
\begin{align*} 
\tilde{d}_k = \sum_{i=1}^M \sum_{t=1}^{k-1} d^\pi_{t,i}~,~
\tilde{U}_k =\sum_{i=1}^M \sum_{t=1}^{k-1} U_{t,i}(d^\pi_{t,i})~,~
\tilde{r}_k = \sum_{t=1}^{k-1} r_{t}, 
\end{align*}
as the sum of exercised consumption, aggregate utility resulting from $\tilde{d}_k$, and aggregate realized renewable generation, respectively. The MPC optimization at $t=k$ is given by
 \begin{equation}
 \begin{array}{lll}
\mathcal{P}^{\mbox{\tiny MPC}}_{\mbox{\tiny NEM X}}: &  \underset{\{d_{t,i}, \forall t\ge k, i\}}{\rm minimize}
&P_{\pi}^{\mbox{\tiny NEM}}\big(\sum_{t=k}^T (\sum_{i=1}^M d_{t,i}-\hat{r}_t) \\&&+ \tilde{d}_k - \tilde{r}_k\big) \\
  && -\sum_{i=1}^M\sum_{t=k}^T U_{t,i}(d_{t,i})-\tilde{U}_k\\
 & \mbox{subject to} & \bar{d}_{t,i} \ge d_{t,i} \ge 0,~~\forall t\geq k, \forall i.\\
 \end{array}
 \end{equation}
We first break the above optimization into three convex optimizations, $\mathcal{P}^{+\mbox{\tiny, MPC}}_{\mbox{\tiny NEM X}}, \mathcal{P}^{-\mbox{\tiny, MPC}}_{\mbox{\tiny NEM X}}$ and $\mathcal{P}^{o\mbox{\tiny, MPC}}_{\mbox{\tiny NEM X}}$, corresponding to the three scheduling zones in:
 \begin{align*}
\begin{array}{lll}
\mathcal{P}^{+\mbox{\tiny, MPC}}_{\mbox{\tiny NEM X}}:& \underset{\{d_{t,i}, \forall t\ge k, i\}}{\rm minimize}&  \pi^+ \big(\tilde{d}_k+\sum_{i=1}^M\sum_{t=k}^T d_{t,i}\\&& - \tilde{r}_k -\sum_{t=k}^T r_t\big) -  \tilde{U}_k\\&&-\sum_{i=1}^M\sum_{t=k}^T U_{t,i}(d_{t,i})\\
 & \mbox{subject to} &  \bar{d}_{t,i} \ge d_{t,i} \ge 0,~~\forall t\geq k, \forall i,\\
 & &   \tilde{d}_k+\sum_{i=1}^M \sum_{t=k}^T d_{t,i},\\&& - \tilde{r}_k- \sum_{t=k}^T r_t  \ge 0. \\
 \end{array}
 \end{align*}
 \begin{align*}
\begin{array}{lll}
\mathcal{P}^{-\mbox{\tiny, MPC}}_{\mbox{\tiny NEM X}}:& \underset{\{d_{t,i}, \forall t\ge k, i\}}{\rm minimize}&  \pi^- \big(\tilde{d}_k+\sum_{i=1}^M\sum_{t=k}^T d_{t,i}\\&& - \tilde{r}_k -\sum_{t=k}^T r_t\big) -  \tilde{U}_k\\&&-\sum_{i=1}^M\sum_{t=k}^T U_{t,i}(d_{t,i})\\
 & \mbox{subject to} &  \bar{d}_{t,i} \ge d_{t,i} \ge 0,~~\forall t\geq k, \forall i,\\
 & &   \tilde{d}_k+\sum_{i=1}^M \sum_{t=k}^T d_{t,i},\\&& - \tilde{r}_k- \sum_{t=k}^T r_t  \leq 0. \\
 \end{array}
 \end{align*}
 \begin{align*}
 \begin{array}{lll}
\mathcal{P}^{o\mbox{\tiny, MPC}}_{\mbox{\tiny NEM X}}:&  \underset{\{d_{t,i}, \forall t\ge k, i\}}{\rm minimize}&   -\tilde{U}_k-\sum_{i=1}^M \sum_{t=k}^T U_{t,i}(d_{t,i})   \\
 & \mbox{subject to} &  \bar{d}_{t,i} \ge d_{t,i} \ge 0,~~\forall t\geq k, \forall i,\\
 & &    \tilde{d}_k+\sum_{i=1}^M \sum_{t=k}^T d_{t,i},\\&& - \tilde{r}_k- \sum_{t=k}^T r_t  = 0. \\
 \end{array}
 \end{align*}
 Given the forecasted $\sum_{t=k}^T r_t$, the optimal schedule is the one that achieves the minimum value among $\mathcal{P}^{+\mbox{\tiny, MPC}}_{\mbox{\tiny NEM X}}, \mathcal{P}^{-\mbox{\tiny, MPC}}_{\mbox{\tiny NEM X}}$ and $\mathcal{P}^{o\mbox{\tiny, MPC}}_{\mbox{\tiny NEM X}}$.

 We prove the Theorem with Lemma~\ref{lem3}-\ref{lem4}.

 \begin{lemma}[Schedule in the net-production and net-consumption zones]  \label{lem3}
 When $\tilde{r}_k+ \sum_{t=k}^T r_t < \tilde{d}_k+\sum_{i=1}^M \sum_{t=k}^T d_{t,i}$, it is optimal to consume with schedule $(d_{t,i}^+), \forall t\geq k,\forall i$.  When $\tilde{r}_k+ \sum_{t=k}^T r_t > \tilde{d}_k+\sum_{i=1}^M \sum_{t=k}^T d_{t,i}$, it is optimal to produce with schedule $(d^-_{t,i}),\forall t\geq k,\forall i$\footnote{In the sequential decision, we only exercise $d_{t,i}^+$ or $d_{t,i}^-$ and then resolve for $k+1$ with realizing the decision and DER production up to and including $k$.}.
 \end{lemma}

 {\em Proof:}    First, we show that, if the prosumer is to consume when $\tilde{r}_k+ \sum_{t=k}^T r_t < \tilde{d}_k+\sum_{i=1}^M \sum_{t=k}^T d_{t,i}$, it is optimal to consume with $(d^+_{t,i}), \forall t\geq k, \forall i$.

 Under $\mathcal{P}^{+\mbox{\tiny, MPC}}_{\mbox{\tiny NEM X}}$, the Lagrangian $\mathcal{L}^+$ is given by
\begin{align*}
\mathcal{L}^+ &= (\pi^+-\mu^+)  \bigg(\tilde{d}_k+\sum_{i=1}^M \sum_{t=k}^T d_{t,i} - \tilde{r}_k- \sum_{t=k}^T r_t \bigg)\\&- \sum_{i=1}^M \sum_{t=k}^T \lambda_{t,i}^+ d_{t,i} +\sum_{i=1}^M \sum_{t=k}^T \gamma_{t,i}^+(d_{t,i}-\bar{d}_{t,i})\\&- \tilde{U}_k-\sum_{i=1}^M \sum_{t=k}^T U_{t,i}(d_{t,i}),\nonumber
\end{align*}
where $\mu^+\geq0$ is the Lagrange multipliers for the net-consumption inequality constraints, and $\lambda^+_{t,i},\gamma^+_{t,i}\geq0$ are the lower and upper limit consumption constraints, respectively. The KKT optimality conditions give that $\forall t\geq k, \forall i$ the optimal schedule $d_{t,i}^+$ and its associated Lagrange multipliers $\mu^+$ and $\lambda^+_{t,i},\gamma^+_{t,i}$ must satisfy
\begin{align*}
V_{t,i}(d_{t,i}^+) = \pi^+-\mu^+ - \lambda_{t,i}^++\gamma^+_{t,i},
\end{align*}
which implies
\[d^+_{t,i} = V_{t,i}^{-1}(\pi^+-\mu^+ - \lambda_{t,i}^++\gamma^+_{t,i}).\]
Because the case $\tilde{d}_k+\sum_{i=1}^M \sum_{t=k}^T d_{t,i} = \tilde{r}_k+ \sum_{t=k}^T r_t$ is covered by $\mathcal{P}^{o\mbox{\tiny, MPC}}_{\mbox{\tiny NEM X}}$, it is without loss of generality to assume that
 \[\tilde{d}_k+\sum_{i=1}^M \sum_{t=k}^T d_{t,i} > \tilde{r}_k+ \sum_{t=k}^T r_t,\] 
which implies that $\mu^+=0$. If $0 \leq V^{-1}_{t,i}(\pi^+) \leq \bar{d}_{t,i}$, then
 \[d_{t,i}^+ = V^{-1}_{t,i}(\pi^+),~~ \lambda_{t,i}^+=\gamma^+_{t,i}=0,\]
 satisfies the part of KKT condition involving device $i$. Therefore, $d_{t,i}^+ = V^{-1}_{t,i}(\pi^+)$ is optimal for device $i$'s consumption. If we have $ V^{-1}_{t,i}(\pi^+)> \bar{d}_{t,i}$, the monotonicity of $V^{-1}$ implies that we can find $d^+_{t,i}=\bar{d}_{t,i}$, with $\lambda_{t,i}^+=0$ and $\gamma^+_{t,i}>0$ satisfying
the KKT condition. Therefore, $d^+_{t,i}=\bar{d}_{t,i}$ is optimal. Likewise, if $V^{-1}_{t,i}(\pi^+)<0$, we must have $d^+_{t,i}=0$.
In summary, the optimal consumption is:
\[d_{t,i}^+ = \max\{0,\min\{ V_{t,i}^{-1}(\pi^+),\bar{d}_{t,i}\}\}, \forall t\geq k, \forall i.\]

\par Next, we show that it is suboptimal to be a net-producer when $\tilde{d}_k+\sum_{i=1}^M \sum_{t=k}^T d_{t,i}^+ > \tilde{r}_k+ \sum_{t=k}^T r_t$.

As a net-producer, the prosumer's schedule is determined by $\mathcal{P}^{-\mbox{\tiny, MPC}}_{\mbox{\tiny NEM X}}$.  The Lagrangian of $\mathcal{P}^{-\mbox{\tiny, MPC}}_{\mbox{\tiny NEM X}}$ is given by
\begin{align*}
\mathcal{L}^- &= (\pi^--\mu^-)  \bigg(\tilde{d}_k+\sum_{i=1}^M \sum_{t=k}^T d_{t,i} - \tilde{r}_k- \sum_{t=k}^T r_t \bigg)\\&  - \sum_{i=1}^M \sum_{t=k}^T \lambda_{t,i}^- d_{t,i} +\sum_{i=1}^M \sum_{t=k}^T \gamma_{t,i}^-(d_{t,i}-\bar{d}_{t,i})\\&-\tilde{U}_k-\sum_{i=1}^M \sum_{t=k}^T U_{t,i}(d_{t,i}),
\end{align*}
where $\mu^-$ is the Lagrange multipliers for the net consumption inequality constraints, and $\lambda^-_{t,i},\gamma^-_{t,i}$ are the lower and upper limit consumption constraints, respectively.  The KKT optimality conditions give that $\forall t\geq k, \forall i$ the optimal schedule $d_{t,i}^-$ and its associated Lagrange multipliers $\mu^-$ and $\lambda^-_{t,i},\gamma^-_{t,i}$ must satisfy
\[
V_{t,i}(d_{t,i}^-) = \pi^-+\mu^- - \lambda_{t,i}^-+\gamma_{t,i}^-.
\label{eq:d_i-}
\]
If the prosumer is a net-energy producer, then $\mu^-=0$, and \[d_{t,i}^-=\max\{0,\min\{ V_{t,i}^{-1}(\pi^+),\bar{d}_{t,i}\}\}, \forall t\geq k, \forall i.\] Therefore, we have
\begin{align*}
\tilde{d}_k+\sum_{i=1}^M \sum_{t=k}^T d_{t,i}^- - \tilde{r}_k+ \sum_{t=k}^T r_t < 0,
\end{align*}
which implies
\begin{align*}
\tilde{r}_k+ \sum_{t=k}^T r_t > \tilde{d}_k+\sum_{i=1}^M \sum_{t=k}^T d_{t,i}^- > \tilde{d}_k+\sum_{i=1}^M \sum_{t=k}^T d_{t,i}^+,\nonumber
\end{align*}
which is a contradiction to $\tilde{r}_k+ \sum_{t=k}^T r_t < \tilde{d}_k+\sum_{i=1}^M \sum_{t=k}^T d_{t,i}^+$.
Finally, the statement that it is optimal for the prosumer to be a net-producer with $(d_{t,i}^-), \forall t\geq k, \forall i$ when $\tilde{r}_k+ \sum_{t=k}^T r_t>\tilde{d}_k+\sum_{i=1}^M \sum_{t=k}^T d_{t,i}^-$ is similarly proved.
\hfill $\blacksquare$

 \begin{lemma}[Schedule in the net-zero zone] \label{lem4}
 When 
 \[\tilde{d}_k+\sum_{i=1}^M \sum_{t=k}^T d_{t,i}^+\le \tilde{r}_k+ \sum_{t=k}^T r_t \le \tilde{d}_k+\sum_{i=1}^M \sum_{t=k}^T d_{t,i}^-,\]
 it is optimal to match the consumption to $\tilde{r}_k+ \sum_{t=k}^T r_t$ with schedule $(d^o_{t,i}(r)), \forall t\geq k, \forall i$ where $d_{t,i}^o(r)$ is continuous and monotonically increasing function of $r$ in $[\tilde{d}_k+\sum_{i=1}^M \sum_{t=k}^T d_{t,i}^+ ~,~ \tilde{d}_k+\sum_{i=1}^M \sum_{t=k}^T d_{t,i}^-]$.
 \end{lemma}

 {\em Proof:} First, we show that, if the prosumer is to be a zero net energy consumer, it is optimal to schedule with $(d_{t,i}^o), \forall t\geq k, \forall i$.

 Under $\mathcal{P}^{o\mbox{\tiny, MPC}}_{\mbox{\tiny NEM X}}$, the Lagrangian is given by
\begin{align*}
\mathcal{L}^o &= \mu^o (\tilde{d}_k+\sum_{i=1}^M \sum_{t=k}^T d_{t,i} - \tilde{r}_k- \sum_{t=k}^T r_t )- \sum_{i=1}^M \sum_{t=k}^T \lambda_{t,i}^o d_{t,i}\\&+\sum_{i=1}^M \sum_{t=k}^T \gamma_{t,i}^o (d_{t,i}-\bar{d}_{t,i})-\tilde{U}_k -\sum_{i=1}^M \sum_{t=k}^T U_{t,i}(d_{t,i}),
\end{align*}
where $\lambda_{t,i}^o,\gamma_{t,i}^o \ge 0, \forall t\geq k, \forall i$.
By the KKT condition, the optimal schedule $d_{t,i}^o, \forall t\geq k, \forall i$ and the associated Lagrange multipliers $\mu^o$ and $\lambda_{t,i}^o,\gamma_{t,i}^o$ must satisfy
\begin{equation*}
V_{t,i}(d_{t,i}^o) = \mu^o - \lambda_{t,i}^o+\gamma_{t,i}^o.
\end{equation*}
Solving the above equation, we have
\[d_{t,i}^o = V_{t,i}^{-1}(\mu^o-\lambda_{t,i}^o+\gamma_{t,i}^o),\]
which similar to Lemma \ref{lem3} gives
\[d_{t,i}^o = \max\{0,\min\{ V_{t,i}^{-1}(\mu^o),\bar{d}_{t,i}\}\}, \forall t\geq k, \forall i,\]

where $\mu^o$ must be such that the equality constraint holds:
\begin{equation}
\tilde{d}_k+\sum_{i=1}^M \sum_{t=k}^T \max\{0, V_{t,i}^{-1}(\mu^o)\} = \tilde{r}_k+\sum_{t=k}^T r_t.
\label{eq:equalityproofAdaptive}
\end{equation}
Next, we show that (\ref{eq:equalityproofAdaptive}) must have a positive solution when $\tilde{d}_k+ \sum_{i=1}^M \sum_{t=k}^T d_{t,i}^+\le \tilde{r}_k+ \sum_{t=k}^T r_t \le \tilde{d}_k+\sum_{t=k}^T d_i^-$,  let
\begin{equation*}
F_k(x):= \tilde{d}_k+\sum_{i=1}^M \sum_{t=k}^T \max\{0, V_{t,i}^{-1}(x)\} -\tilde{r}_k-\sum_{t=k}^T r_t.
\end{equation*}
Note that $F_k(\cdot)$ is continuous and monotonically decreasing.  Because
 \begin{equation*}
 F_k(\pi^+) \le 0,~~F_k(\pi^-) \ge 0,
 \end{equation*}
there must exists $\mu^o \in [\pi^-,\pi^+]$ such that $F_k(\mu^o)=0$. Therefore, (\ref{eq:equalityproofAdaptive}) must have positive solution, which also implies that for every $t\geq k, i$:
\[
d_{t,i}^+ \le d_{t,i}^o(r) \le d^-_{t,i}.
\]
Furthermore, the continuity and monotonicity of $F_k$ in $r$  implies that $d_{t,i}^o(r), \forall t\geq k, \forall i$ is continuous and monotonically increasing function of $r$.
Finally, we show that,  when $\tilde{d}_k+\sum_{i=1}^M \sum_{t=k}^T d_{t,i}^+ \le \tilde{r}_k+ \sum_{t=k}^T r_t \le \tilde{d}_k+\sum_{i=1}^M \sum_{t=k}^T d_{t,i}^-$,  it is suboptimal to net-consume or net-produce.

Consider again $\mathcal{P}^{+\mbox{\tiny, MPC}}_{\mbox{\tiny NEM X}}$ when the prosumer is to consume optimally.  From the solution of $\mathcal{P}^{+\mbox{\tiny, MPC}}_{\mbox{\tiny NEM X}}$, we have
\begin{align*}
\tilde{d}_k+\sum_{i=1}^M \sum_{t=k}^T d_{t,i}^+ > \tilde{r}_k+ \sum_{t=k}^T r_t,
\end{align*}
which is a contradiction. The case that it is suboptimal to net-produce is similarly proved.  \hfill $\blacksquare$
\subsubsection{Sequential consumption decision algorithm}
An implementation of the MPC-based sequential consumption decision is illustrated in Algorithm~\ref{alg:ConsumptionMPC}.

\begin{algorithm}
\caption{Sequential consumption decision}\label{alg:ConsumptionMPC}
\begin{algorithmic}[0]
\State \textbf{Input:} tariff parameter $\pi$, marginal utility for every $i$ device $V_{t,i}\in \mathbb{R}^T$, $r$ measurements, and consumption limits $\bar{d}_{t,i}$
  \State \textbf{Output:} optimal consumption decision
\State \textbf{Initialize:} $\tilde{d}_0 \gets 0, \tilde{r}_0 \gets 0$
\For{$k \gets 1,T$}
\State $d^+_k:= \tilde{d}_k+\sum_{i=1}^M \sum_{t=k}^T d_{t,i}^+$ 
\State $d^-_k:= \tilde{d}_k+\sum_{i=1}^M \sum_{t=k}^T d_{t,i}^-$
\State $\tilde{r}_k \gets \sum_{t=1}^{k-1} r_t$
\State \textbf{Forecast} $\hat{r}: k \rightarrow T$
\For{\textbf{all} device $i \in \mathcal{M}$}
\If{$\tilde{r}_k+\sum_{t=k}^T \hat{r}_t \leq  d^+_k$}
    \State $d^\pi_{t,i} \gets d^+_{t,i}$
\ElsIf{$\tilde{r}_k+\sum_{t=k}^T \hat{r}_t \in \left[d^+_k , d^-_k\right]$}
    \State Solve for $\mu$ \[\tilde{d}_k+\sum_{i=1}^M \sum_{t=k}^T \max{(0,\min{(V^{-1}_{t,i}(\mu),\bar{d}_{t,i})})} = \tilde{r}_k+\sum_{t=k}^T  \hat{r}_t\]
    \State Compute the net-zero zone consumption $\bm{d}_i^o(\bm{r})$
    \State $\bm{d}_i^o(\bm{r}) = \max\{\bm{0}, \min \{V_i^{-1}(\bm{1}\mu^*(\bm{r})),\bar{\bm{d}}_i\}\} \in \mathbb{R}^T$
    \State $d^\pi_{t,i} \gets d^o_{t,i}$
    \Else
    \State $d^\pi_{t,i} \gets d^-_{t,i}$
\EndIf
\EndFor
\EndFor
\State The optimal consumption vector is $\bm{d}^\pi_i$ for all $i \in \mathcal{M}$ 
\end{algorithmic}
\end{algorithm}

\subsection{Extra numerical results}
\label{sec:AppendixMiscResults}
\subsubsection{Short-run: NEM X active and passive prosumers}
\begin{figure}[htbp]
    \centering
    \includegraphics[scale=0.32]{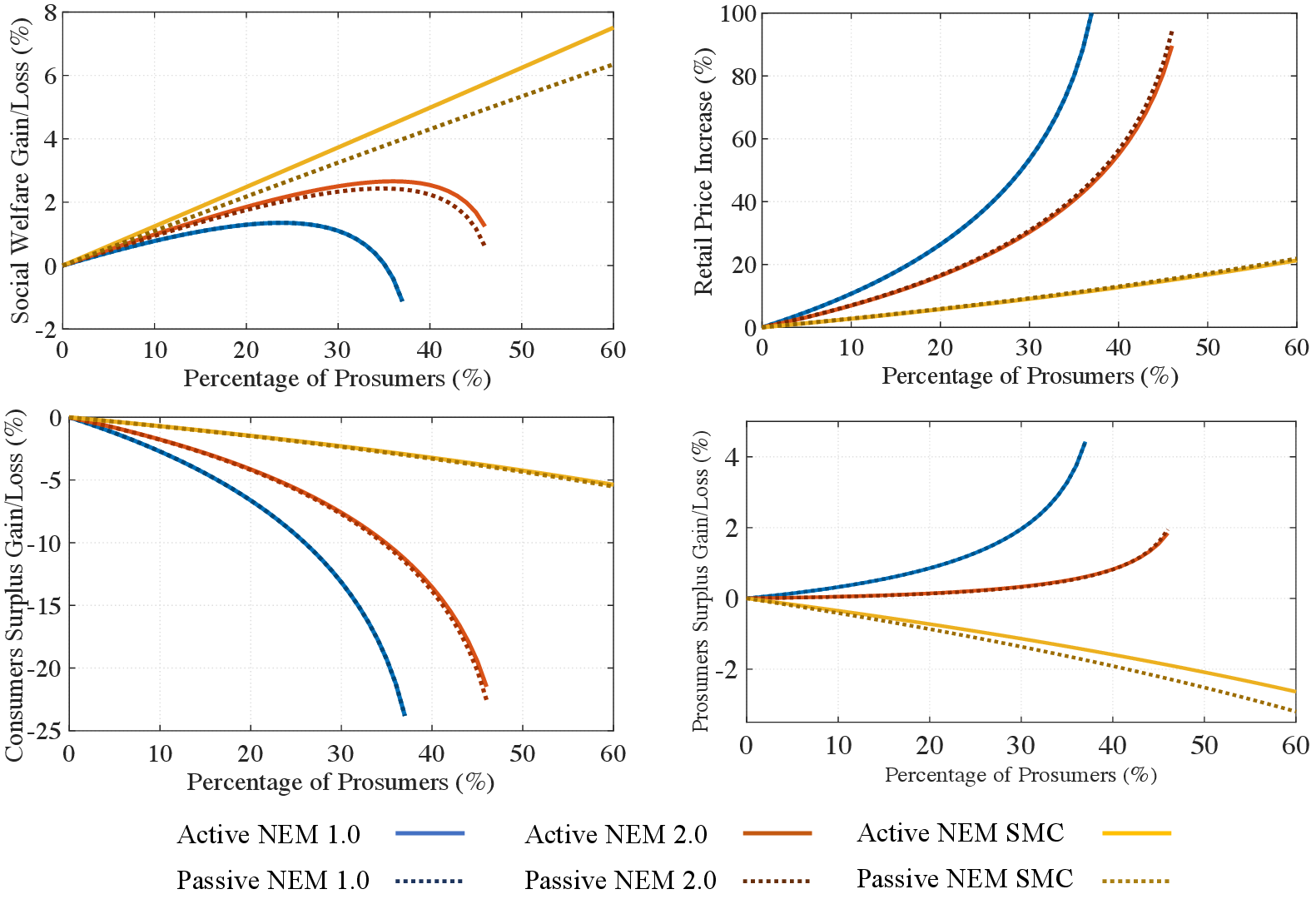}
    \caption{Active and passive NEM X. Clockwise from top-left: social welfare, retail price, prosumer surplus, and consumer surplus percentage changes.}
    \label{fig:ActivePassiveNum}
\end{figure}

Figure \ref{fig:ActivePassiveNum} shows the percentage change (over the no adoption case $\gamma=0$) in the total social welfare, the retail prices, and the active and passive NEM X prosumers and consumers surpluses as functions of the prosumer population size $\gamma$.
\par under both active and passive NEM 1.0 cases, the social welfare was equivalent. The reason was that under NEM 1.0, the optimal consumption of the active prosumer is independent of $r$, which makes it equivalent to the passive prosumer's optimal consumption. Once $\pi^+>\pi^-$, the retail price increase (top-right panel of Fig.\ref{fig:ActivePassiveNum}) under the passive prosumer was faster because the passive prosumer pays less toward the utility company (and therefore gets more monetary compensation) due to the lower consumption ($d^+ \leq d^-$). When the prosumer pays less toward the utility, the revenue adequacy constraint is more strained, which drove the retail rate to increase faster under the passive prosumer scenario. As a result of the faster price increase under the passive prosumers case when $\pi^+>\pi^-$, the consumer surplus (bottom-left panel of Fig.\ref{fig:ActivePassiveNum}) decayed faster under the passive prosumer case.

\par The prosumer surplus (bottom-right panel of Fig.\ref{fig:ActivePassiveNum}) explanation is more involved. In the 2.0 case, which has a small price differential, the passive prosumer surplus increased at a rate slower than the active prosumer surplus. The behavior shifted when the retail rate increase became significant. The reason for this behavior is that as the retail rate increased (which means that the sell rate also increased), the percentage difference between $\pi^+$ and $\pi^-$ became smaller, which means that the surplus due to increasing the consumption (from $d^+$ to $d^-$) became also smaller. In other words, net exportation value increased, while self-consumption value decreased, therefore, the passive prosumers surplus catches up when the retail and sell rates simultaneously increased. This simultaneous increase, however, does not hold under the SMC policy, which resulted in a faster decay of the passive prosumer surplus for the whole trajectory.

\par Lastly, the expected social welfare percentage change (top-left panel of Fig.\ref{fig:ActivePassiveNum}) under the active prosumer case (solid lines) is always higher than the welfare under the passive case (dashed lines). This is because under the passive prosumers, the consumer surplus increase is slower, and the prosumer surplus decrease is faster. Figure.\ref{fig:ActivePassiveNum} overall, signals the importance of enabling DER adopters to exercise DER-elastic consumption decisions.

\begin{figure}[htbp]
    \centering
    \includegraphics[scale=0.22]{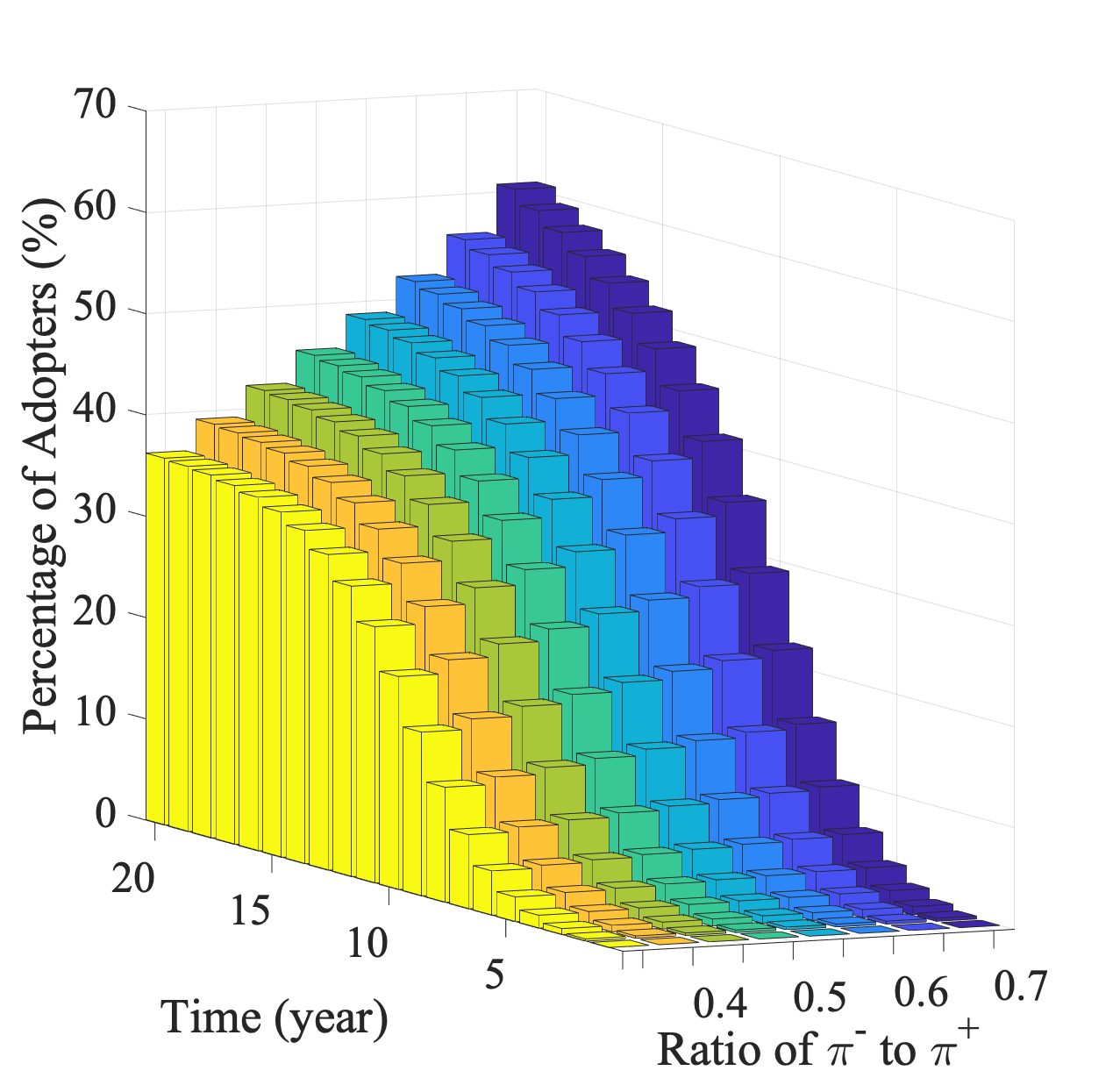}
    \caption{Long-run adoption under different compensation rates.}
    \label{fig:adoption_sellratexx}
\end{figure}
\begin{figure}[htbp]
    \centering
    \includegraphics[scale=0.22]{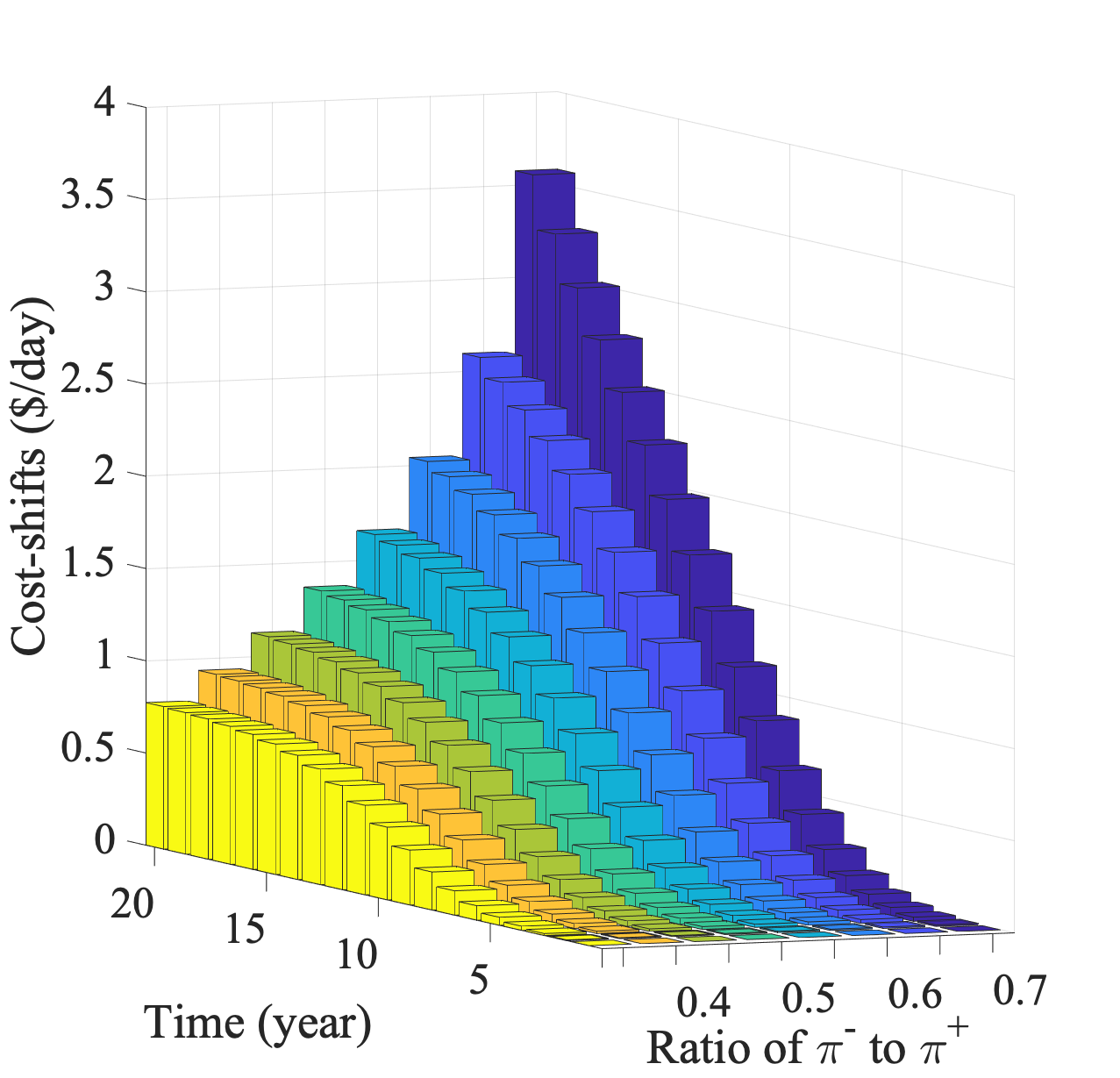}
    \caption{Long-run cost-shifts under different compensation rates.}
    \label{fig:costshift_sellrate}
\end{figure}
\begin{figure}[htbp]
    \centering
    \includegraphics[scale=0.25]{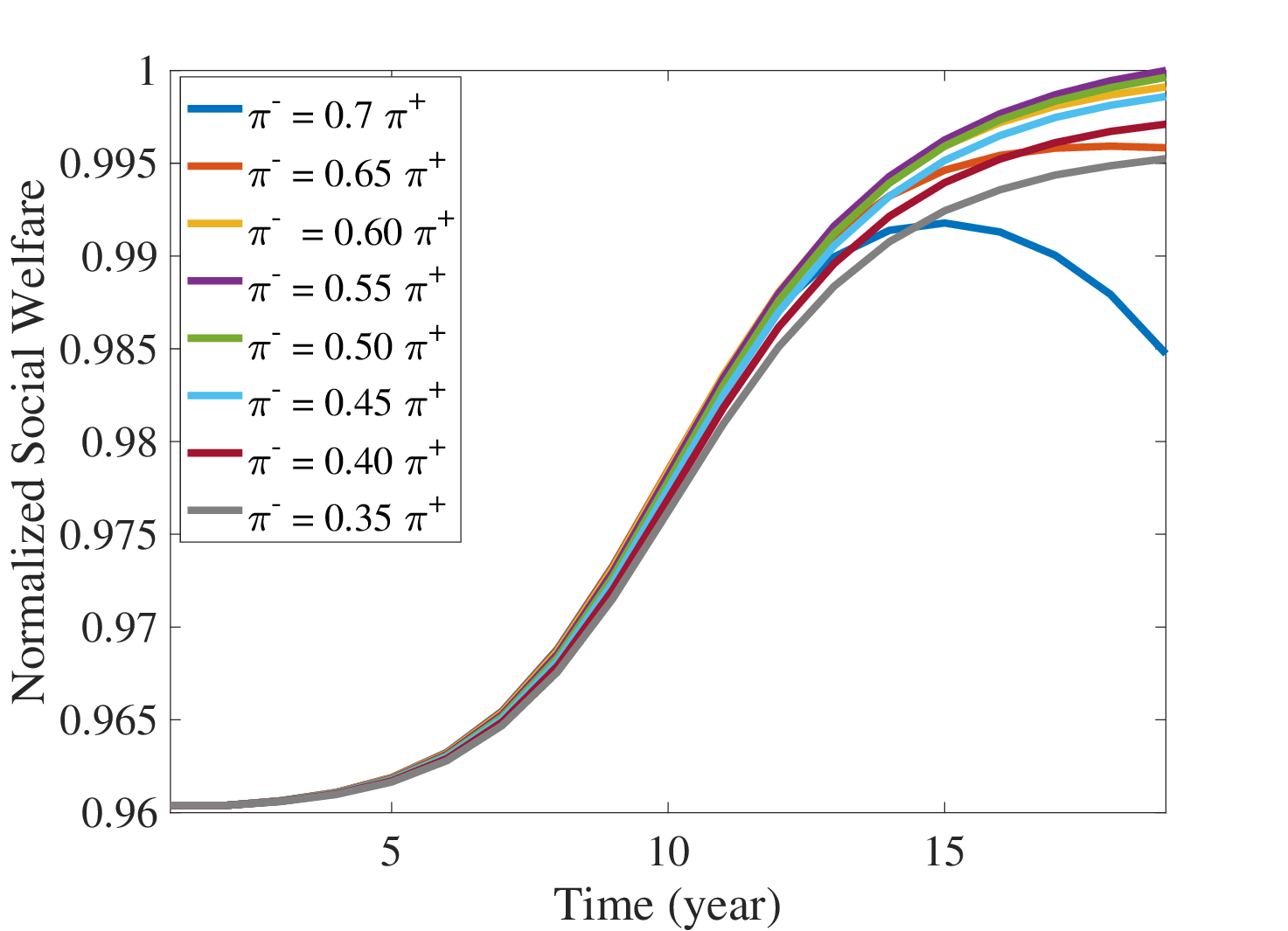}
    \caption{Long-run social welfare under different compensation rates.}
    \label{fig:welfare_sellratexx}
\end{figure}

\subsubsection{Long-run: Export rate effect}

Figures \ref{fig:adoption_sellratexx}-\ref{fig:welfare_sellratexx} explored the long-run effect of export compensation rate on market adoption (Fig.\ref{fig:adoption_sellratexx}), cross-subsidies (Fig.\ref{fig:costshift_sellrate}) and social welfare (Fig.\ref{fig:welfare_sellratexx}). The exogenous parameters $\theta$ and $\xi$ were fixed, but over the evolution of states, we assumed the average PV installation cost to be only 30\% of the initial (current average installation cost, which was $\xi_0=$ \$4500/kW\footnote{The average 2019 solar cost data for systems less than 10kW in California can be found at:\url{https://www.californiadgstats.ca.gov/charts/nem}.}. 
\par The long-run adoption curves of 8 compensation rates ranging from 0.35\% to 0.7\% of the retail rate $\pi^+$ with an increment of 0.5\% are shown in Fig.\ref{fig:adoption_sellratexx}. Higher compensation rates such as $\pi^- = 0.7\pi^+$ ushered rooftop solar adoption, but at the cost of higher subsidies and lower social welfare as shown in Fig.\ref{fig:costshift_sellrate}-\ref{fig:welfare_sellratexx}. Reducing this compensation rate to $0.35\pi^+$ shrunk the percentage of adopters to less than 40\% of the market customers, which shows how can the sole change of compensation rates effectively influence the adoption decisions and the diffusion in the long-run. The monotonically increasing adoption under all compensation rates is driven by the monotonicity of the retail rate with the fraction of adopters.  

\par The cost-shifts resulting from each compensation policy (Fig.\ref{fig:costshift_sellrate}) proved that lower compensation rates reduce price markups between $\pi^+,\pi^-$ and $\pi^{\mbox{\tiny SMC}}$, resulting in lower bill savings, that are closer to the utility's avoided cost due to BTM generation. Fig.\ref{fig:costshift_sellrate} shows that a policy that compensates excess solar, for example, at 70\% of the retail rate yielded an averaged cost-shift that is 69\% higher than a policy that compensates at 30\% of the retail rate.
\par Lastly, Fig.\ref{fig:welfare_sellratexx} showed the implicit inter-play between the surplus of adopters and non-adopters as the adoption process evolves under the different compensation rates. Whereas lower compensation rates yielded higher consumers' surplus and lower prosumers' surplus, since the retail rate was lower, higher compensation rates yielded higher prosumers' surplus, which comes primarily at the cost of consumers' surplus. Therefore, the relatively very high ($\pi^-=0.7\pi^+$) and very low ($\pi^-=0.35\pi^+$) compensation rates yielded the lowest social welfare. An intermediate compensation rate at $0.55\pi^+$ gave the highest social welfare since it mildly compensates excess solar giving the optimal compromise between consumer surplus, prosumer surplus, and environmental benefits.

\subsection{Numerical results data}
\label{sec:appendixNumData}
In this section, we describe the data sources of the numerical results. The solar data profile is taken from the California Solar Initiative (CSI) 15-Minute interval PV data\footnote{which can be found at: {\url{https://www.californiadgstats.ca.gov/downloads}}}. The solar PV cost in California in 2019 is $\xi_0 = 4500 \$/kW$\footnote{The average 2019 solar cost data for systems less than 10kW in California can be found at: \url{https://www.californiadgstats.ca.gov/charts/nem}.}. Since the date is from California, we used the utility's fixed cost  $\theta_{\mbox{\tiny PGE}}$ of PG\&E using publicly available revenue, MWh sales, and the number of customers data of Pacific Gas and Electric Company (PG\&E)\footnote{Revenue, sales, and the number of customers of PG\&E data was taken from EIA over the years from 2016-2019: \url{https://www.eia.gov/electricity/data/state/} .}. The value was $\theta_{\mbox{\tiny PGE}} = \$2.86$/customer/day. The SMC rate was assumed to be the sum of the LMP rate\footnote{The day-ahead LMP data is taken from CAISO SP15 for the period June-August, 2019. The data can be found at: \url{http://oasis.caiso.com/mrioasis/logon.do}.} $\pi^\omega$ and the non-market cost of pollution that was reflected on the retail price \cite{Next10Report}\footnote{The non-market cost of pollution was calculated based on the avoided non-energy cost due to BTM DER, which was estimated by \cite{Borenstein_avoidedCost:20Haas} to be \$0.012/kWh, and the renewable portfolio standard (RPS) compliance benefits estimated to be \$0.018/kWh, as mentioned in: \url{https://www.sce.com/regulatory/tariff-books/rates-pricing-choices/renewable-energy-credit}.}. To compute the environmental benefits $\mathcal{E}$ in (\ref{eq:Ramseymaximization}), which has the form in \cite{Alahmed_Tong:22IEEETSG}, the price $\pi^e$ was quantified at \$0.035/kWh solar from \cite{Wiser_EnvrionmentalVoS:16LBNL}.

\par We adopt a widely-used quadratic concave utility function of the form:
\begin{equation}
    \label{eq:example-utility-function}
    U_i(d_i) = \alpha_i d_i - \frac{1}{2}\beta_i d_i^2, 
\end{equation}
where $\alpha_i, \beta_i$ are some utility parameters that are dynamically calibrated.
\par Three load types with three different utility functions of the form in (\ref{eq:example-utility-function}) were considered: 1) HVAC load\footnote{\label{HVAC} The residential load profile data is taken from NREL open dataset for a nominal household in Los Angeles. We used the summer months data, that is June-August, 2019. The data can be found at: \url{https://shorturl.at/uyL36}.}, 2) EV load\footnote{The EV load data is taken from NREL EV Infrastructure Projection (EVI-Pro) simulation tool for the city of Los Angeles, CA: \url{https://afdc.energy.gov/evi-pro-lite/load-profile}}, 3) other household loads such as lighting and appliances\textsuperscript{\ref{HVAC}}.
As introduced in \cite{campaigne_balandat_ratliff_2016}, the historical retail prices\footnote{We use historical PG\&E prices, which can be found at: \url{https://www.pge.com/tariffs/electric.shtml}.} and historical consumption data are used to calibrate the quadratic utility function parameters by predicating an elasticity of demand\footnote{The HVAC and household appliances elasticity values are taken from \cite{ASADINEJAD_Elasticity:18EPSR}, and the EV charging elasticity value is taken from \cite{Tong_EVmarket:17JAERE}}. Considering that for historical data, the price differential is zero, then an interior solution of the prosumer problem in section \ref{sec:prosumer} yields:
\[ d_i^h(\pi^h) = \frac{\alpha_i-\pi^h}{\beta_i},
\]
where $\pi^h$ is the historical retail price. For each load type $i$ having an elasticity of demand $\varepsilon_i$, the elasticity can be expressed as:
\[
\varepsilon_i (\pi^h) = \frac{\partial d_i^h(\pi^h)}{\partial \pi^h} \frac{\pi^h}{d_i^h}  =- \frac{1}{\beta_i}\frac{\pi^h}{d_i^h} =-\frac{\pi^h}{\alpha_i-\pi^h} .
\]
Solving for $\alpha_i$ and $\beta_i$, we get:
\[\alpha_i = -  \left( \frac{1-\varepsilon_i}{\varepsilon_i}\right)\pi^h \]
\[\beta_i = -\frac{\pi^h}{\varepsilon_i d_i^h}.\]
$\alpha_i$ and $\beta_i$ are calibrated for each time period based on the realized prices and consumption data.

\end{document}